\UseRawInputEncoding
\documentclass[11pt,draftcls,onecolumn]{IEEEtran}

\ifCLASSOPTIONcompsoc
  \usepackage[nocompress]{cite}
\else
  \usepackage{cite}
\fi

\usepackage{enumerate}
\usepackage{latexsym}
\usepackage{amsfonts}
\usepackage{amsmath}
\usepackage{amssymb}
\usepackage{color}
\usepackage{colortbl}
\usepackage{epsfig}
\usepackage{xspace}
\usepackage{graphicx}
\usepackage{arcs}

\usepackage{subfigure}
\usepackage{balance}
\usepackage[english]{babel}

\ifCLASSINFOpdf
\else
\fi
\usepackage{epsfig}
\usepackage{multirow}
\usepackage{url}

\newtheorem{lemma}{Lemma}
\newtheorem{theorem}{Theorem}
\newtheorem{conclusion}{Conclusion}
\newtheorem{proposition}{Proposition}
\newtheorem{remark}{Remark}

\newtheorem{corollary}{Corollary}
\newtheorem{claim}{Claim}
\newenvironment{proof}{
        \vspace{1ex}
        {\noindent\bf Proof:}}{\eop\vspace{1ex}}
\newcommand{\eop}{\hspace*{\fill}\mbox{$\Box$}}     

\begin{document}

\title{Asymptotic  Critical Transmission Radii in Wireless Networks over a Convex Region
\thanks{A earlier version of this work was submitted  to \emph{IEEE Transactions on Information Theory} on Nov. 3rd, 2018, as the first round submission with Ref. IT-18-0767. The revised version
was submitted to \emph{IEEE Transactions on Information Theory} on Dec. 18th, 2021, as a second round submission with Ref. IT-21-0894.
}
}

\author{Jie Ding, 
        Shuai Ma,
       and Xin-Shan Zhu
\IEEEcompsocitemizethanks{\IEEEcompsocthanksitem J. Ding   is with China Institute of FTZ Supply Chain, Shanghai Maritime University, Shanghai 201306, China.
\protect\\
E-mail: jieding78@hotmail.com (Correspondence Author).


\IEEEcompsocthanksitem  S. Ma  is with SKLSDE Lab,
Beihang University, Beijing 100191, China\protect\\
E-mail: mashuai@buaa.edu.cn.

\IEEEcompsocthanksitem X. Zhu is with   School of Electrical and Information Engineering, Tianjin University, Tianjin 300072, China.\protect\\
E-mail: xszhu126@126.com (Correspondence Author).

}


}

\markboth{IEEE Transactions on Information Theory,~Vol.~X, No.~X, Dec. 2021}%
{Ding et al.: Asymptotic  Critical Transmission Radii in Wireless Networks over a Convex Region}

\IEEEtitleabstractindextext{%
\begin{abstract}
Critical transmission ranges (or radii) in wireless ad-hoc and sensor networks have been extensively investigated for
various performance metrics such as connectivity, coverage, power assignment and energy consumption. However, the regions on which the networks are
distributed are typically either squares or disks in existing works, which seriously limits the usage in real-life applications.
In this article, we consider a convex region  (i.e., a generalisation of squares and disks) on which wireless nodes are uniformly distributed.
We have investigated two types of critical transmission radii, defined in terms of
$k-$connectivity and the minimum vertex degree, respectively, and have also established their precise asymptotic distributions.
These make the previous results obtained under the circumstance of squares or disks special cases of this work. More importantly,
our results  reveal how the region shape impacts on the critical transmission ranges:  it is the length of the boundary
of the (fixed-area) region that completely determines the transmission ranges. Furthermore, by isodiametric inequality, the smallest critical transmission ranges  are achieved when regions are disks only.
\end{abstract}

}

\maketitle

\begin{abstract}
Critical transmission ranges (or radii) in wireless ad-hoc and sensor networks have been extensively investigated for
various performance metrics such as connectivity, coverage, power assignment and energy consumption. However, the regions on which the networks are
distributed are typically either squares or disks in existing works, which seriously limits the usage in real-life applications.
In this article, we consider a convex region  (i.e., a generalisation of squares and disks) on which wireless nodes are uniformly distributed.
We have investigated two types of critical transmission radii, defined in terms of
$k-$connectivity and the minimum vertex degree, respectively, and have also established their precise asymptotic distributions.
These make the previous results obtained under the circumstance of squares or disks special cases of this work. More importantly,
our results  reveal how the region shape impacts on the critical transmission ranges:  it is the length of the boundary
of the (fixed-area) region that completely determines the transmission ranges. Furthermore, by isodiametric inequality, the smallest critical transmission ranges  are achieved when regions are disks only.
\end{abstract}


\section{Introduction}
\label{sec-intro}

\par Let $\chi_n$ be a uniform $n$-point process over a
unit-area convex region $\Omega$, i.e., a set of $n$ independent
points each of which is uniformly distributed over $\Omega$, and every pair of points whose Euclidean distance less than $r_n$ is
connected with an undirected edge. So  a random geometric
graph $G(\chi_n,r_n)$ is obtained.
Random geometric graphs
are a basic model of wireless sensor and ad-hoc networks,  and have drawn significant
attention~\cite{Penrose-RGG-book, DC02:PR-E-random-geometric-graphs, Santi05:Critical-Transmitting-Range-Mobile-AdHoc, ZMMa07:Coverage-WSN, HWang14:Critical-sensing-camera, XBWwang12:Coverage-Energy-Consumption-Control, FKKumar10:Fundamentals-Connectivity-Capacity, PJWan09:Critical-Radii-Greedy-Routing, PWan11:Asymptotic-distribution-Greedy, JWan05:Power-Assignment-kConnectivity, XBWang13:Mobility-Increases-Connectivity, PWan04:Asymptotic-critical-transmission-MobileHoc, PJWan-IT-asymptotic-radius}.


 $k-$connectivity and
the smallest vertex degree are two interesting topological
properties of a random geometry graph. A graph $G$ is said to be
$k-$connected if there is no set of $k-1$ vertices whose removal
would disconnect the graph. Denote by $\kappa$ the connectivity of
$G$, being the maximum $k$ such that $G$ is $k-$connected.
The minimum vertex degree of $G$ is denoted by $\delta$.   Let
$\rho(\chi_n;\kappa\geq k)$ be the minimum $r_n$ such that
$G(\chi_n,r_n)$ is $k-$connected and $\rho(\chi_n;\delta\geq k)$ be
the minimum $r_n$ such that $G(\chi_n,r_n)$ has the smallest degree $k$, respectively.

%
%
%
%
%
%
%
%
%
%

It is known that various network performance metrics, such as coverage, capacity,
throughput and energy consumption, depend on the critical transmission range (or radius) defined in terms of
$\rho(\chi_n;\kappa\geq k)$ and $\rho(\chi_n;\delta\geq k)$. Indeed, critical transmission or sensing ranges have been extensively investigated, e.g.,
 for connectivity in mobile ad-hoc networks~\cite{Santi05:Critical-Transmitting-Range-Mobile-AdHoc},
 for coverage in wireless sensor networks~\cite{ZMMa07:Coverage-WSN} and in mobile heterogeneous camera sensor networks~\cite{XBWwang12:Coverage-Energy-Consumption-Control,HWang14:Critical-sensing-camera},
 for capacity in large scale sensor networks~\cite{FKKumar10:Fundamentals-Connectivity-Capacity},
 for routing in wireless ad-hoc networks~\cite{PJWan09:Critical-Radii-Greedy-Routing,PWan11:Asymptotic-distribution-Greedy},
 for  power assignment in wireless ad-hoc networks~\cite{JWan05:Power-Assignment-kConnectivity}, and for energy consumption control in mobile heterogeneous wireless sensor networks~\cite{XBWwang12:Coverage-Energy-Consumption-Control}.
Further, the effect of mobility on the critical transmission range for asymptotic connectivity in $k-$hop clustered networks has also been studied~\cite{XBWang13:Mobility-Increases-Connectivity}.

However, the regions on which the networks are distributed, considered in all the above mentioned works, are usually  unit-area squares or disks.
Here the area of considered regions is always assumed to be unit one. It has been indicated by Wan \emph{et al.} in ~\cite{PWan04:Asymptotic-critical-transmission-MobileHoc,PJWan-IT-asymptotic-radius}, that the  shape of the  region has impacts on the critical transmission radii $\rho(\chi_n;\kappa\geq k)$ and $\rho(\chi_n;\delta\geq k)$,  by comparing their precise asymptotic distributions on squares and disk regions.
In realistic scenarios, the network nodes are in general placed in a more general region rather than a simple square or disk. Therefore, the assumption of square or disk regions limits the application of those existing works on the analysis of critical transmission ranges~\cite{Santi05:Critical-Transmitting-Range-Mobile-AdHoc, ZMMa07:Coverage-WSN, HWang14:Critical-sensing-camera, XBWwang12:Coverage-Energy-Consumption-Control, FKKumar10:Fundamentals-Connectivity-Capacity, PJWan09:Critical-Radii-Greedy-Routing, PWan11:Asymptotic-distribution-Greedy,XBWang13:Mobility-Increases-Connectivity, PWan04:Asymptotic-critical-transmission-MobileHoc,PJWan-IT-asymptotic-radius}.

In this article, we instead consider a more general unit-area convex region $\Omega$, on which $n$ wireless sensors are uniformly distributed. The critical transmission or sensing ranges $\rho(\chi_n;\kappa\geq k)$ and $\rho(\chi_n;\delta\geq k)$ over the unit-area convex region are investigated. In particular, we will establish the precise asymptotic distributions of these two types of radii, which makes the previous results obtained under the circumstance of squares or disks special cases of our work.
More importantly, our results will reveal how the region shape impacts on the critical transmission ranges. That is, it is only the length of the boundary
of the region that determines the critical transmission ranges. Furthermore, by isodiametric inequality, the smallest asymptotic critical transmission radii  are achieved only when the region $\Omega$ is a disk. 


 \begin{table*}
  \begin{center}\caption{Notations and descriptions}\label{table:notation}
       \begin{tabular}{|c|c|}
         \hline\hline
         Notations & Descriptions \\ \hline
         $\Omega$ & An unit-area convex region on $\mathbb{R}^2$ \\\hline
          $\partial\Omega$ &  The boundary of region $\Omega$ \\\hline
          $\chi_n$ &   A uniform $n$-point process over $\Omega$   \\\hline
          $ \mathcal{P}_n$ &  A homogeneous Poisson process of intensity $n$ (i.e., $n|\Omega|$) on $\Omega$  \\\hline
         $G(\chi_n,r_n)$ & A graph obtained by connecting each pair of points in $\chi_n$ if there distance
         is less than $r_n$   \\\hline

          $\rho(\chi_n;\kappa\geq k)$  &   The minimum $r_n$ such that
$G(\chi_n,r_n)$ is $k-$connected   \\\hline
          $\rho(\chi_n;\delta\geq k)$ &  The minimum $r_n$ such that $G(\chi_n,r_n)$ has the smallest degree $k$         \\\hline
           $\rho(\mathcal{P}_n;\delta\geq k)$& The minimum $r_n$ such that $G( \mathcal{P}_n,r_n)$ has the smallest degree $k$ \\\hline
           $G(\Omega)$ &  An uniform upper bound of all
second-order derivatives\\
   &  of  curve $\partial \Omega$ in local coordinates (see Remark~\ref{remark1})  \\\hline
          $B(x,r)$   &  A ball centered at $x$ with radius $r$ \\\hline
           $|\cdot|$ & Area of a measurable set on $\mathbb{R}^2$  \\\hline
            $\|\cdot\|$   & Length of a line segment \\\hline
            $ \mathrm{dist}(x,\partial \Omega)$
             & $ \mathrm{dist}(x,\partial \Omega)=\inf_{y\in \partial\Omega}\|xy\|$, the distance of point $x$ to curve $\partial \Omega$. \\\hline
            $ \psi^k_{n,r}(x)$ & Short for $\frac{1}{k!}(n|B(x,r)\cap \Omega|)^k\exp(-n|B(x,r)\cap \Omega|)$  \\\hline
            $a(r,t)$ (or shortly, $a(t)$) & Denotes the area $|\{x=(x_1,x_2):x_1^2+x_2^2\leq r^2, x_1\leq t\}|$ \\\hline
            $\Pr(\cdot), \Pr[\cdot], \Pr\{\cdot\}$ & The probability of an event \\\hline
            $f(n)=\Theta(g(n))$   &   $c_1g(n)\leq f(n)\leq c_2g(n)$ for sufficiently large $n$ where $0<c_1<c_2$    \\\hline
            $f(n)=o(g(n))$ &   $\lim\limits_{n\rightarrow\infty}\frac{f(n)}{g(n)}=0$ \\\hline
            $f(n)\sim g(n)$ &    $\lim\limits_{n\rightarrow\infty}\frac{f(n)}{g(n)}=1$   \\\hline
       \end{tabular}
    \end{center}
    \end{table*}

\par
We use the following notations throughout this article.
(1) Region $\Omega \subset \mathbb{R}^2$ is a unit-area convex  region on a plane, and $B(x,r)\subset \mathbb{R}^2$ is a ball
centered at $x$ with  radius $r$.  A bounded set  means that this set can be included in a ball with a finite radius.
(2) Notation $|A|$ is a shorthand for the
2-dimensional Lebesgue measure (i.e., area) of a measurable set $A\subset \mathbb{R}^2$. All integrals considered will be Lebesgue integrals.
(3) Notation $\|\cdot\|$ represents the length of a line segment. For instance, $\|BC\|$ is the length of line segment $BC$.
(4) Given any two nonnegative functions $f(n)$ and $g(n)$, if there exist two constants $0<c_1<c_2$ such that
$c_1g(n)\leq f(n)\leq c_2g(n)$ for sufficiently large $n$, then we denote $f(n)=\Theta(g(n))$.
We also use notations $f(n)=o(g(n))$ and $f(n)\sim g(n)$ to denote that $\lim\limits_{n\rightarrow\infty}\frac{f(n)}{g(n)}=0$ and $\lim\limits_{n\rightarrow\infty}\frac{f(n)}{g(n)}=1$, respectively. { For convenience, the notations with descriptions are presented in Table~\ref{table:notation}, some of which will be introduced later.}

The remainder of this article is structured as follows.
We first present our main results and related work in Section~\ref{sec-related-mainresults}.
Preliminary results, i.e. some lemmas, are provided in Section~\ref{sec-lemmas}.
The proof of the main result is divided into two parts, which are demonstrated in Sections~\ref{sec:proof-part1} and~\ref{sec:proof-part2} respectively. We finally conclude this article in Section~\ref{sec-conclusion}.

\section{Main Results and Related Work}
\label{sec-related-mainresults}

\par There are quite a few studies  on the precise asymptotic distribution of the
critical transmission radius in terms of $\rho(\chi_n;\delta\geq k)$ and $\rho(\chi_n;\kappa\geq k)$.
In this section, after reviewing closely related work, we shall provide our main results on the precise
asymptotic distributions.

\subsection{Related Work}

When regions $\Omega$ are squares, the precise
asymptotic distribution of $\rho(\chi_n;\delta\geq 1)$ has been
derived by Dette and Henze in~\cite{DH-largest-NN} as follows,
\begin{equation}\label{eq:distribution-rho-delta-1}
     \Pr\left\{\rho(\chi_n;\delta\geq 1)\leq r_n\right\}\sim \exp\left(-e^{-c}\right),
\end{equation}
where  $c$ is a positive constant and
$$
r_n=\sqrt{\frac{\log n+c}{\pi n}}.
$$
When regions $\Omega$ is a disk and $k$-connectivity is concerned, this $r_n$ also satisfies the following
\begin{equation}\label{eq:distribution-rho-connectivity-1}
    \Pr\left\{\rho(\chi_n;\kappa\geq 1)\leq r_n\right\}\sim \exp\left(-e^{-c}\right),
\end{equation}
which gives the precise
asymptotic distribution of $\rho(\chi_n;\kappa\geq 1)$. This was given by Gupta and Kumar in~\cite{GK-Critical-Power}.
It has been proved by Penrose in~\cite{Penrose97:longest-edge} that this critical transmission radius for connectivity equals to the longest edge of the minimal spanning tree, and he has further provided the asymptotic distribution of the longest edge in~\cite{Penrose99:Strong-law-longest-edge}.
In addition, a type of $(k,m)-$connectivity was defined by Wang in his monograph~\cite{XBWang2013:MotionCast-book}, together with the precise asymptotic distribution of the critical transmission range over a square region.

For a square region $\Omega$,
Penrose has further demonstrated the following result:
\begin{theorem}\label{thm:Penrose-formulae}(\cite{Penrose-RGG-book,Penrose-k-connectivity})
 Let $k>0,\lambda\in \mathbb{R}^+$, $\Omega$ be a unit-area
square.
 If sequence $\left(r_n\right)_{n\geq 1}$ satisfies
     \begin{equation}\label{eq:formula}
        \frac n{k!}\int_{\Omega}\left(n|B(x,r_n)\cap\Omega|\right)^k e^{-n|B(x,r_n)\cap\Omega|}dx\sim \lambda,
     \end{equation}
then as $n\rightarrow\infty$, the probabilities of the two events
$\rho(\chi_n;\delta\geq k+1)\leq r_n$ and $\rho(\chi_n;\kappa\geq
k+1)\leq r_n$ both converge to $e^{-\lambda}$.
\end{theorem}
\par
The formula~(\ref{eq:formula}) is derived through    Poisson approximation and de-Poissonized techniques,  exploited by Penrose in the proof of this theorem.
{ Let $\mathcal{P}_n$ be a homogeneous Poisson process of intensity $n$ (i.e., $n|\Omega|$) on $\Omega$, then $\mathcal{P}_n (\Omega)=\sum_{i=1}^m\mathcal{P}_n(A_i)$
where $\Omega=\cup_{i=1}^mA_i$ is a union of disjoint of $A_i\subset\Omega~(i=1,2,\cdots,m)$, $\mathcal{P}_n (\Omega)$ is a Poisson random variable with intensity $n|\Omega|$, and $\mathcal{P}_n(A_i)\ (i=1,2,\cdots,m)$ are mutually independent Poisson random variables with intensities $n|A_i|\ (i=1,2,\cdots,m)$ respectively.}
 That is, the number of the points in $A_i$, i.e., $\mathcal{P}_n(A_i)$, satisfies the following
Poisson distribution:
$$
   \Pr\left\{\mathcal{P}_n(A_i)=j\right\}=\frac{e^{-n|A_i|}(n|A_i|)^j}{j!},\quad j\geq0.
$$
Given the condition that there are exact $j$ points in any subregion $A\subset\Omega$, then these $j$ points are independently and uniformly distributed in $A$. This fact makes $\chi_n$ can be well approximated by $\mathcal{P}_n$.

Here Penrose assumed that $\Omega$ was a square. In~\cite{PWan04:Asymptotic-critical-transmission-MobileHoc} and \cite{PJWan-IT-asymptotic-radius}, Wan \emph{et
al.} have indicated, without a proof,  that this theorem also holds for a unit-area disk.
Furthermore, they  have  provided the explicit forms of $r_n$ satisfying    formula
(\ref{eq:formula}) when $\Omega$ is a square and disk, respectively. See the following two theorems.
\begin{theorem}(\cite{PWan04:Asymptotic-critical-transmission-MobileHoc, PJWan-IT-asymptotic-radius})
Assume that $\Omega\subset \mathbb{R}^2$ is a unit-area square. Let
     $$
        r_n=\sqrt{\frac{\log n+(2k-1)\log\log n+\xi}{\pi n }},
     $$
where $\xi$ satisfies
   $$ \xi=\left\{\begin{array}{cc}
      -2\log \left(\sqrt{e^{-c}+\frac{\pi}{4}}-\frac{\sqrt{\pi}}{2}\right), & k=1,\\
      2\log\frac{\sqrt{\pi}}{2^{k-1}k!}+2c, & k>1, \\
    \end{array}
        \right.
$$
then   formula (\ref{eq:formula}) holds.
\end{theorem}

\begin{theorem}(\cite{PWan04:Asymptotic-critical-transmission-MobileHoc, PJWan-IT-asymptotic-radius})
Assume that $\Omega\subset \mathbb{R}^2$ is a unit-area disk. Let
     $$
        r_n=\sqrt{\frac{\log n+(2k-1)\log\log n+\xi}{\pi n }},
     $$
where $\xi$ satisfies
   $$ \xi=\left\{\begin{array}{cc}
      -2\log \left(\sqrt{e^{-c}+\frac{\pi^2}{16}}-\frac{\pi}{4}\right), & k=1, \\
      2\log\frac{{\pi}}{2^{k}k!}+2c, & k>1. \\
    \end{array}
        \right.
$$
then   formula~(\ref{eq:formula}) holds.
\end{theorem}

\par As long as   formula~(\ref{eq:formula}) is satisfied,
the probabilities of the two events
$\rho(\chi_n;\delta\geq k+1)\leq r_n$ and $\rho(\chi_n;\kappa\geq
k+1)\leq r_n$ both converge to $e^{-\lambda}$ as $n$ tends to
infinity, followed by Theorem~\ref{thm:Penrose-formulae}.

However, in realistic scenarios the regions on which
networks are distributed, are typically more general than disks and squares. Unfortunately, to the best of our knowledge, there are no
results on the topic of the critical transmission radius, obtained under the circumstance of
a more general region, e.g. a convex region.

\subsection{Our Main Results}
\label{subsec-mainresults}

Whether Theorem~\ref{thm:Penrose-formulae} holds for a general
convex region remains an open problem (for more details,
please see page 176 in~\cite{Penrose-RGG-book}). However,
Penrose has pointed out that, without a proof in his monograph
\cite{Penrose-RGG-book}, Theorem~\ref{thm:Penrose-formulae} should also hold for the case
of polyhedral domain. The  contribution of this article is to extend  the conclusion of Theorem~1 to a general
convex region case,
with additionally providing an explicit form of $r_n$.
See the following Theorem~\ref{thm:combined-Main}.
\begin{theorem}\label{thm:combined-Main}
Let $\Omega\subset \mathbb{R}^2$ be a unit-area convex region such that the length of
the boundary $\partial\Omega$  is $l$, $k\geq 0$ be an integer and $c>0$ be a constant.
\par (i) If $k>0$, let
     \begin{equation}\label{eq:Theorem-formula-1}
        r_n=\sqrt{\frac{\log n+(2k-1)\log\log n+\xi}{\pi n }},
     \end{equation}
where $\xi$ satisfies
$$ \left\{\begin{array}{cc}
     \xi=-2\log \left(\sqrt{e^{-c}+\frac{\pi l^2}{64}}-\frac{l\sqrt{\pi}}{8}\right), & k=1, \\
      \xi=2\log \left(\frac{l\sqrt{\pi}}{2^{k+1}k!}\right)+2c, & k>1. \\
    \end{array}
        \right.
$$
\par (ii) If $k=0$, let
     \begin{equation}\label{eq:Theorem-formula-2}
        r_n=\sqrt{\frac{\log n+c}{\pi n }}.
     \end{equation}
Then
    \begin{equation}\label{eq:Theorem-formula-3}
        \frac n{k!}\int_{\Omega}\left(n|B(x,r_n)\cap\Omega|\right)^k e^{-n|B(x,r_n)\cap\Omega|}dx\sim e^{-c},
    \end{equation}
and therefore,  the probabilities of the two events
$\rho(\chi_n;\delta\geq k+1)\leq r_n$ and $\rho(\chi_n;\kappa\geq
k+1)\leq r_n$ both converge to $\exp\left(-e^{-c}\right)$  as $n\rightarrow\infty$.
\end{theorem}

This theorem  generalise the previous conclusions  presented in Theorems~1, Theorem~2 and Theorem~3, and makes them special cases.
In particular, in Theorem~\ref{thm:combined-Main}, if the region $\Omega$ is a unit-area square (disk, respectively), then $l=4$ ($l=2\sqrt{\pi}$, respectively), and then the determined
 $r_n$ coincides with the one presented in Theorem~2 (Theorem~3, respectively) with a simple calculation.

In \cite{PWan04:Asymptotic-critical-transmission-MobileHoc, PJWan-IT-asymptotic-radius}, Wan \emph{et al}. mentioned that in
Theorems~2 and~3, ``$\xi$ depends on the shape of $\Omega$ and parameter $k$''.  Our conclusions, i.e.,
Theorem~\ref{thm:combined-Main}, clearly indicates  that the asymptotic radius is completely and only determined by
 the length $l$ of the boundary $\partial\Omega$ and the parameter $k$. These
  findings have never been revealed before,  to the best of our knowledge.
In comparison, the claim of Wan \emph{et al.} is rather rough because regions with different shapes may have the same perimeters. In addition,
 only two specific regions are examined in their work.

Furthermore, Theorem~\ref{thm:combined-Main} shows that smaller the perimeter is,
the smaller the asymptotic critical radius is.
 By the following isoperimetric inequality on $\mathbb{R}^2$~\cite{Bandle:Isoperimetric-inequalities-book}:
 $$
      l^2\geq 4\pi S(\Omega)=4\pi,
 $$
where $S(\Omega)=1$ is the area of $\Omega$, and the equality only hods for a disk, we know that
a disk has the smallest perimeter among all equal-area regions.
From this, we have the following

\begin{corollary}
The smallest asymptotic critical radius is achieved only when the region $\Omega$ is a disk.
\end{corollary}

It is worth pointing out that when $k=0$, i.e., when one minimum vertex degree or $1-$connectivity is concerned, the shape of the region has no impacts
on the radius. This conclusion also coincides with the results  (\ref{eq:distribution-rho-delta-1}) and (\ref{eq:distribution-rho-connectivity-1}),
given by  Dette \emph{et al.} and  Gupta \emph{et al.} respectively.

\begin{remark}
However, if energy saving for networks is taken into account, and the minimum degree is greater than two or $k-$connectivity ($k\geq 2$) is considered, the shape that the distributed network nodes formed should be like a ``disk'', because the disk shape can lead to the smallest critical transmission ranges. Otherwise, for example, suppose $n$ nodes are placed uniformly in a rectangle with width $\epsilon$ and length $\frac{1}{\epsilon}$, the critical transmission range must be quite large to guarantee the network to be $k-$connected ($k\geq2$), if $\epsilon$ is small (i.e., the perimeter $2\epsilon+\frac{2}{\epsilon}$ is large). Therefore, our theoretical results have a practical meaning as well.
\end{remark}

%
%

{
The proof of Theorem~\ref{thm:combined-Main} will be divided into two parts. The first part is to prove that $r_n$ given by
(\ref{eq:Theorem-formula-1}) and (\ref{eq:Theorem-formula-2}) satisfies
$$
        \frac n{k!}\int_{\Omega}\left(n|B(x,r_n)\cap\Omega|\right)^k e^{-n|B(x,r_n)\cap\Omega|}dx\sim e^{-c},
$$
which will be  demonstrated in Section~\ref{sec:proof-part1}. The second part, i.e., the probabilities of the two events
$\rho(\chi_n;\delta\geq k+1)\leq r_n$ and $\rho(\chi_n;\kappa\geq
k+1)\leq r_n$ both converge to $\exp\left(-e^{-c}\right)$  as $n\rightarrow\infty$, will be proved in Section~\ref{sec:proof-part2}. At first, we give some preliminary results in the following section.}

\section{Preliminary results}
\label{sec-lemmas}

\begin{lemma}\label{lemma:LowerBound-B(x,r)} Let $\Omega$  be a bounded convex region,
then there exists a positive constant $C$ such that for any
sufficiently small $r$,
$$
        \mathop{\inf}_{x\in \Omega}|B(x,r)\cap \Omega |\geq C\pi
        r^2.
$$
\end{lemma}
\begin{proof}
See Appendix~A.
\end{proof}

\begin{figure}[ht]
\centering
\includegraphics[width=5cm]{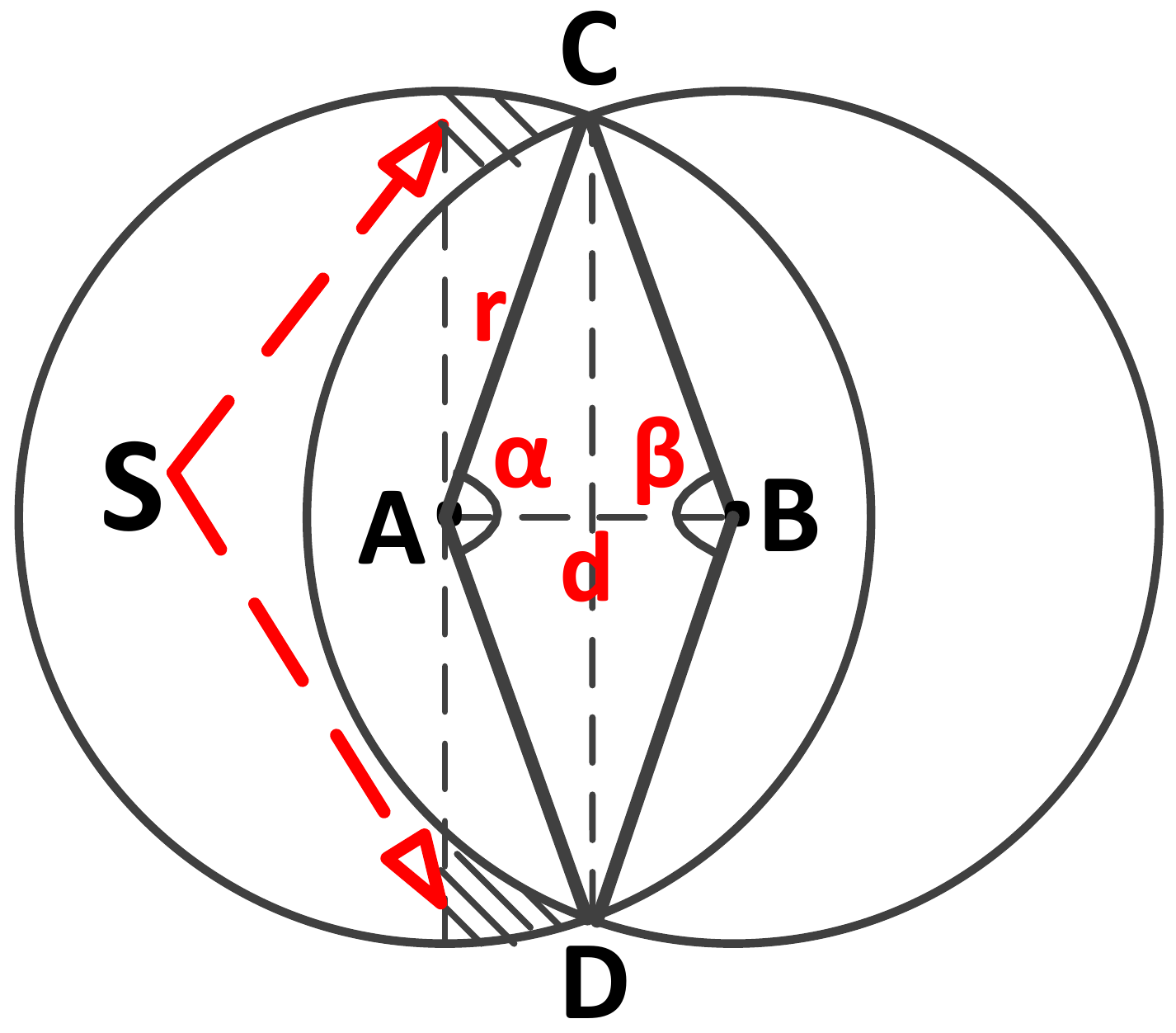}
\caption{Illustration of Lemma~\ref{lem:shadow-low-bound}: area $S$ }
\label{fig:Lemma2-Figure5-1}
\end{figure}

\begin{lemma}\label{lem:shadow-low-bound}
Two disks centered at $A$ and $B$ respectively, have  the same radii $r$, with the distance $\|AB\|=d$.
The shadow area $S$ in
Figure~\ref{fig:Lemma2-Figure5-1} is
$$S=r^2 \left[\frac{1}{2}\frac{d}{r} + \frac{19}{48} \left(\frac{d}{r}\right)^3 + \frac{153}{1280}\left(\frac{d}{r}\right)^5
- o\left(\frac{d}{r}\right)^5\right].$$
In particular, if $r\geq d\geq \frac{r}{(nr^2)^{\frac12}}=\frac{1}{\sqrt{n}}$, then
$$nS\geq \frac{1}{\sqrt{\pi}}\left(n\pi r^2\right)^{\frac12}.$$
\end{lemma}
\begin{proof}
See Appendix~A.
\end{proof}

\par The next lemma needs the condition of smooth boundary of $\Omega$. Throughout this article, the boundary $\partial \Omega$  is smooth means that the  function of curve $\partial \Omega$ on $\mathbb{R}^2$  has second-order continuous derivatives everywhere.

Now we assume  region $\Omega$ has a smooth boundary $\partial \Omega$.  Let $A,B$ be any two points on $\partial \Omega$. Consider the
$X-$axis in a coordinate system, which is parallel to line $AB$ and tangent to it at point $O$.
Notice that the slope of the curve $\partial \Omega$ increases from
a negative value at point $B$ to a positive value at point $A$, and is zero at
point $O$. We also find a rectangle
$ABCD$, whose side $CD$ is on $X-$axis and tangent to $\partial \Omega$ at point
$O$. See Figure~\ref{fig:Rectangle-ABCD}. The rectangle $ABCD$
satisfies the following property:

 \begin{figure}[htbp]
 \begin{center}
       \scalebox{0.50}[0.50]{
          \includegraphics{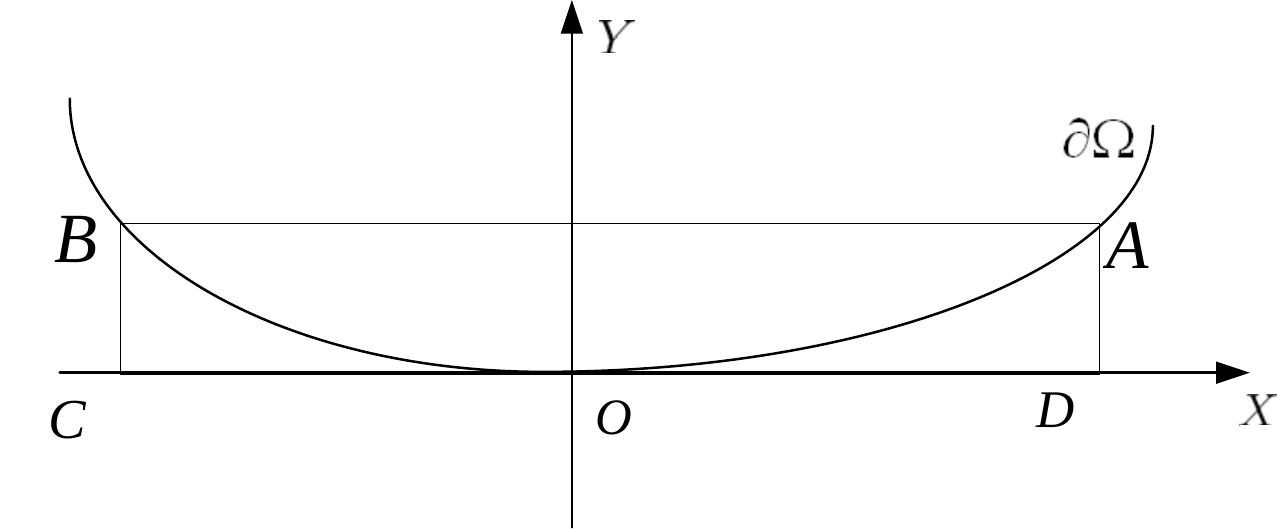}
       }\caption{$\partial \Omega$ is tangent to $CD$ at point $O$}\label{fig:Rectangle-ABCD}
  \end{center}
  \end{figure}

\begin{lemma}\label{lemma:RectangleArea}
Denote $h$ by the length of  line segment $AB$. Then $S_{ABCD}$, the area of
 rectangle $ABCD$, satisfies $S_{ABCD}=\Theta(h^3)$ or
$S_{ABCD}=o(h^3)$, as $h\rightarrow 0$.
\end{lemma}

\begin{proof} By the Taylor expansion of  $\partial \Omega$'s function
at tangent point $O$, we obtain
  \begin{eqnarray}
  \|AD\|&=& \mathrm{constant}\cdot \|OD\|^2+o(\|OD\|^2)
    \end{eqnarray}
where $\|\cdot\|$ is the length of a line segment as mentioned, and
$constant$ is the second-order derivative of the function of $\partial \Omega$ at
point $O$. Here the first-order derivative of $\partial \Omega$'s equation at point $O$ is zero in this coordinate system. Notice that
 $\|OD\|\leq\|AB\|=h$, so
$$S_{ABCD}=\|AB\|\cdot\|AD\|\leq \mathrm{constant}\cdot h^3+o(h^3).$$
The lemma is proved.
\end{proof}

\begin{remark}\label{remark1}
By the smoothness of the bounded and closed curve $\partial \Omega$, we may
assume that all this kind of $constant$s, i.e., all
second-order derivatives of $\partial \Omega$ at all tangent
points in all such local coordinate systems, have a uniform upper
bound $G(\Omega)>0$.  Therefore, as long as $h$ is sufficiently small we will have that
\begin{equation}
      \|OD\|\geq \sqrt{\frac{\|AD\|}{G(\Omega)+1}},
\end{equation}
\begin{equation}
 \|OC\|\geq  \sqrt{\frac{\|BC\|}{G(\Omega)+1}}=\sqrt{\frac{\|AD\|}{G(\Omega)+1}},
\end{equation}
which leads to that if $\|AD\|> (G(\Omega)+1)r^2$, then
\begin{equation}\label{eq:Remark-equation}
\|OD\|> r, \quad \|OC\|> r.
\end{equation}
This
conclusion will also be used in the proof of
Theorem~\ref{thm:combined-Main}.
\end{remark}

\begin{figure}[htbp]
 \begin{center}
       \scalebox{0.81}[0.8]{
                    \includegraphics{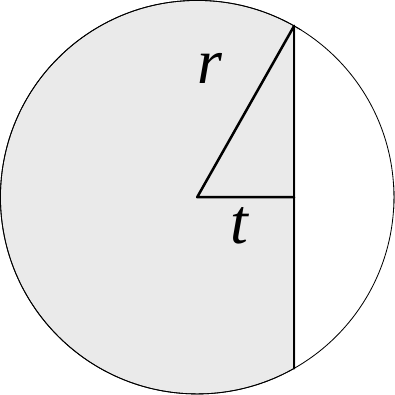}
       }
      \caption{$a(t)$ is the area of the shade region (see~\cite{PWan04:Asymptotic-critical-transmission-MobileHoc})}\label{fig:a(t)}
  \end{center}
  \end{figure}

The following lemma is first given by Wan \emph{et al.} in~\cite{PWan04:Asymptotic-critical-transmission-MobileHoc}. For any $t\in[0,r]$, we
define
\begin{equation}
  a(r,t)=|\{x=(x_1,x_2):x_1^2+x_2^2\leq r^2, x_1\leq t\}|,
\end{equation}
the area of the shaded region illustrated in Figure~\ref{fig:a(t)}.
We may shortly write it as $a(t)$, then
\begin{lemma}\label{lemma:a(t)}(\cite{PWan04:Asymptotic-critical-transmission-MobileHoc})
If $r=r_n=\sqrt{\frac{\log n+(2k-1)\log\log n+\xi}{\pi n }}$, then
\begin{enumerate}
  \item If $k>0$,
$$
    n\int_0^{\frac r2}\frac{(na(t))^ke^{-na(t)}}{k!}dt\sim
    \frac{\sqrt{\pi}}{2^{k+1}k!}e^{-\frac{\xi}2}.
$$
  \item If $k=0$,
$$
  n\int_0^{\frac r2}\frac{(na(t))^ke^{-na(t)}}{k!}dt=o(1).
$$
\end{enumerate}
\end{lemma}
This lemma and its proof can be found in~\cite{PWan04:Asymptotic-critical-transmission-MobileHoc} or~\cite{PJWan-IT-asymptotic-radius}.
For convenience, as we have mentioned, throughout this
article we sometimes use $r$ to represent $r_n$. We also assume that $r$ is sufficiently small and $n$ is sufficiently large, if necessarily.

\section{Proof of Theorem~\ref{thm:combined-Main}: part one}\label{sec:proof-part1}
\label{sec-theorem4}

\par In this section, we will present a complete proof for the first part of Theorem~\ref{thm:combined-Main}:

\par (i) If $k>0$, let
     $$
        r_n=\sqrt{\frac{\log n+(2k-1)\log\log n+\xi}{\pi n }},
     $$
where $\xi$ satisfies
$$ \left\{\begin{array}{cc}
     \xi=-2\log \left(\sqrt{e^{-c}+\frac{\pi l^2}{64}}-\frac{l\sqrt{\pi}}{8}\right), & k=1, \\
      \xi=2\log \left(\frac{l\sqrt{\pi}}{2^{k+1}k!}\right)+2c, & k>1. \\
    \end{array}
        \right.
$$
\par (ii) If $k=0$, let
     $$
        r_n=\sqrt{\frac{\log n+c}{\pi n }}.
     $$
Then
     \begin{equation}\label{eq:proof-formulae}
        \frac n{k!}\int_{\Omega}\left(n|B(x,r_n)\cap\Omega|\right)^k e^{-n|B(x,r_n)\cap\Omega|}dx\sim e^{-c}.
     \end{equation}

{
The following estimation about $r_n$ given in the theorem is straight forward, and will be often used.  We list them as a remark.
\begin{remark}\label{rem:remark3}
$r_n$ given in Theorem~\ref{thm:combined-Main} satisfies
\begin{enumerate}

   \item $ {n\pi r^2}\rightarrow\infty$ as $n\rightarrow\infty$.

  \item Let $\xi$ be the one given in Theorem~\ref{thm:combined-Main}, then
  \begin{eqnarray*}
     \frac n{k!}(n\pi r^2)^k e^{-n\pi r^2}\sim \left\{
              \begin{array}{cc}
                e^{-\xi}, & k=1, \\
                 o(1), & k>1. \\
              \end{array}
            \right.
\end{eqnarray*}
  \item  If $C>0$ is a constant, then
 $$
     (n\pi r^2)^{k+1}\exp(-Cn\pi r^2)\sim \frac{(\log n+o(\log n))^{k+1}}{e^{C(\log n+o(\log n))}}=o(1).
$$
\end{enumerate}
\end{remark}

Throughout this article, we define
\begin{equation}\label{eq:psi-function}
       \psi^k_{n,r}(x)= \frac{\left(n|B(x,r)\cap\Omega|\right)^k
       e^{-n|B(x,r)\cap\Omega|}}{k!}.
\end{equation}
All left work  is to prove
     \begin{equation}\label{eq:proof-formulae-2}
        n\int_{\Omega} \psi^k_{n,r}(x)dx\sim e^{-c}.
     \end{equation}
}

The pioneering work by Wan \emph{et al.}
in~\cite{PWan04:Asymptotic-critical-transmission-MobileHoc} and ~\cite{PJWan-IT-asymptotic-radius}, i.e.,  the methods of coping with square and disk regions  in the proofs of Theorem~2
and~3, stimulates us to generalise them into
a uniform framework to deal with any general convex region.
The proof of the above part  of Theorem~\ref{thm:combined-Main} is divided into two cases. The first case is to assume the boundary $\partial\Omega$ to be smooth.
In this case, region $\Omega$ is divided into four disjoint parts, i.e.,\ $\Omega=\Omega(0)\cup\Omega(2)\cup\Omega(1,2)\cup\Omega(1,1)$. The proof
in this case will thus be completed within three steps: Step~1 deals with the integrals on $\Omega(0)$ and $\Omega(2)$, Step~2 copes with $\Omega(1,2)$
while Step~3 with $\Omega(1,1)$. After that, we further treat the continuous boundary case through an approximation approach.

\subsection{Case of smooth $\partial \Omega$}
In this subsection, we assume the boundary $\partial \Omega$ is smooth.
Let
$$
     \Omega(0)=\left\{x\in\Omega:\mathrm{dist}(x,\partial\Omega)\geq r\right\},
$$
$$
     \Omega(2)=\left\{x\in\Omega:\mathrm{dist}(x,\partial\Omega)\leq
     (G(\Omega)+1)r^2\right\},
$$
$$
     \Omega(1)=\Omega\backslash\left(\Omega(0)\cup\Omega(2)\right).
$$
Here the constant $G(\Omega)$ is given by Remark~\ref{remark1}, and $\mathrm{dist}(x,\partial\Omega)$, as we mentioned, represents the distance between point $x$ and curve $\partial\Omega$. { Here $r$ is considered sufficiently small (in fact, $r\rightarrow 0$  as $n\rightarrow\infty$). Therefore,
$(G(\Omega)+1)r^2<r$, and thus $\Omega(0)$ and $\Omega(2)$ are disjoint.
}
Clearly, we have
\begin{equation*}
\begin{split}
&n\int_{\Omega}\psi^k_{n,r}(x)dx\mathrm{d}x=n\left(\int_{\Omega(0)}+\int_{\Omega(2)}+\int_{\Omega(1)}\right)\psi^k_{n,r}(x)dx\mathrm{d}x.
\end{split}
\end{equation*}

The integrations of $n\psi^k_{n,r}(x)$ on these subregions, i.e., $\Omega(0), \Omega(2), \Omega(1)$, will be estimated respectively. The first two estimates are rather simple, see the following Proposition~\ref{pro:Omega(0)} and Proposition~\ref{pro:Omega(2)}.
The estimation of $n\int_{\Omega(1)} \psi^k_{n,r}(x)dx$ will be demonstrated in Proposition~\ref{pro:Omega(1,2)} and Proposition~\ref{pro:Omega(1,1)}.

\begin{proposition}\label{pro:Omega(0)}  The integration of $ \psi^k_{n,r}(x)$ defined by (\ref{eq:psi-function}) on subregion $\Omega(0)$ satisfies that
$$n\int_{\Omega(0)} \psi^k_{n,r}(x)dx\sim \left\{
              \begin{array}{cc}
                e^{-\xi}, & k=1, \\
                 o(1), & k>1. \\
              \end{array}
            \right.$$
\end{proposition}
\begin{proof}
$\forall x\in \Omega(0), |B(x,r)\cap\Omega|=\pi r^2$. { Notice that $|\Omega(0)|\sim (1-lr)$ where $l$ is the length of
$\partial \Omega$. So, by Remark~\ref{rem:remark3},}
\begin{eqnarray*}
     n\int_{\Omega(0)} \psi^k_{n,r}(x)dx
     &=&\frac n{k!}(n\pi r^2)^k e^{-n\pi r^2}|\Omega(0)|\\
     &\sim &\frac n{k!}(n\pi r^2)^k e^{-n\pi r^2}\\
     &\sim &\left\{
              \begin{array}{cc}
                e^{-\xi}, & k=1, \\
                 o(1), & k>1. \\
              \end{array}
            \right.
\end{eqnarray*}
\end{proof}

\begin{proposition}\label{pro:Omega(2)}The integration of $ \psi^k_{n,r}(x)$ defined by (\ref{eq:psi-function}) on subregion $\Omega(2)$ satisfies that
\begin{eqnarray*}
   n\int_{\Omega(2)} \psi^k_{n,r}(x)dx=o(1).
\end{eqnarray*}
\end{proposition}
\begin{proof} By Lemma~\ref{lemma:LowerBound-B(x,r)}, there exists $C>0$ such
that for any $x\in \Omega$, we have $|B(x,r)\cap\Omega|\geq C\pi
r^2$. Notice that $$|\Omega(2)|\leq l\times (G(\Omega)+1)r^2=\Theta(1)r^2,$$
where $l$ is the length of the boundary $\partial\Omega$.  Then by Remark~\ref{rem:remark3},
\begin{eqnarray*}
   n\int_{\Omega(2)} \psi^k_{n,r}(x)dx&\leq &\frac n{k!}(n\pi r^2)^k e^{-Cn\pi r^2}|\Omega(2)|\\
     &=&\Theta(1)(n\pi r^2)^{k+1} e^{-Cn\pi r^2}\\
     &=&o(1).
\end{eqnarray*}
\end{proof}

\begin{figure}[htbp]
 \begin{center}
       \scalebox{0.7}[0.7]{
           \includegraphics{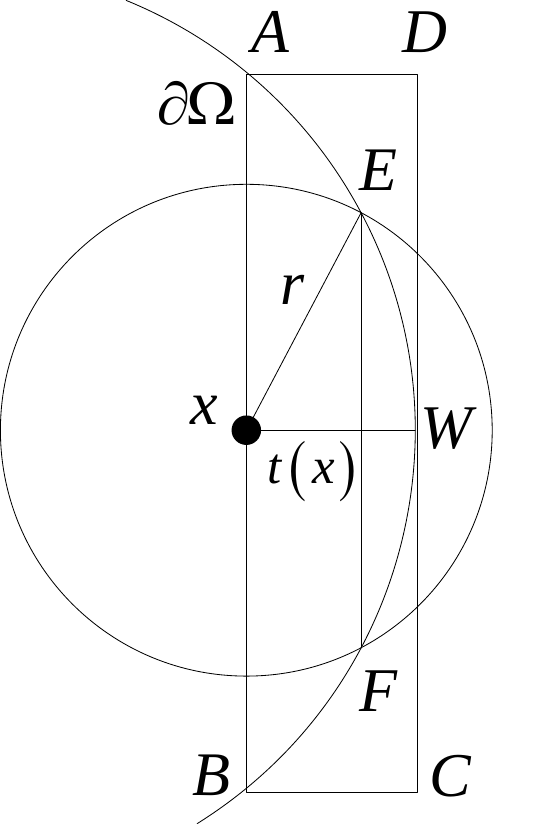}       }
       \caption{$t(x)$ is nonnegative}\label{positive-a(t)}
  \end{center}
  \end{figure}

Now we consider the integration of $n\psi^k_{n,r}(x)$ on  subregion
$\Omega(1)$. 
If $x\in \Omega(1)$, then
\begin{equation}\label{eq:Omega(1)-distance}
(G(\Omega)+1)r^2<\mathrm{dist}(x,\partial\Omega)<r,
\end{equation}
 and thus
the circle centred at $x$ with radius $r$ intersects $\partial
\Omega$ at two points, namely $E$ and $F$ (see
Figure~\ref{positive-a(t)} or Figure~\ref{fig:UpperBound-B(x,r)Omega}). The distance
from $x$ to  chord $EF$ of the circle is defined by $t(x,r)$ or
shortly $t(x)$. Furthermore, we  set
$$
    \Omega(1,1)=\left\{x\in \Omega(1):t(x)\leq \frac r2\right\},
$$
$$
   \Omega(1,2)=\Omega(1)\setminus\Omega(1,1),
$$
then
\begin{equation*}
n\int_{\Omega(1)}\psi^k_{n,r}(x)dx=n\left(\int_{\Omega(1,1)}+\int_{\Omega(1,2)}\right)\psi^k_{n,r}(x)dx.
\end{equation*}

\begin{figure}[htbp]
 \begin{center}
       \scalebox{0.7}[0.7]{
           \includegraphics{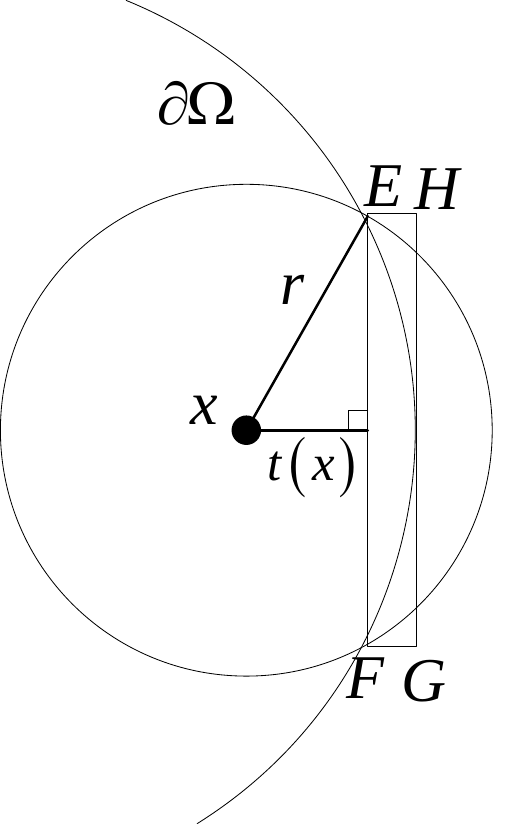}       }
       \caption{Upper bound for $|B(x,r)\cap
       \Omega|$}\label{fig:UpperBound-B(x,r)Omega}
  \end{center}
  \end{figure}

In following, we will specify $n\int_{\Omega(1,2)}\psi^k_{n,r}(x)dx$ in
Proposition~\ref{pro:Omega(1,2)}, and then  determine $n\int_{\Omega(1,1)}\psi^k_{n,r}(x)dx$ in
Proposition~\ref{pro:Omega(1,1)}.
But at first, we will establish a lower and upper bound for $|B(x,r)\cap\Omega|$ where $x\in\Omega(1)$, which will be used in the proof of
these two propositions.

For point $x\in \Omega(1)$, let $W\in\partial \Omega$ be the nearest point to $x$, i.e.,
$$
  \|xW\|=\mathrm{dist}(x,\partial \Omega).
$$
Let line $CD$ be tangent to $\partial \Omega$ at point $W$,  and  line
$AB$, which parallels $CD$ and passes through point $x$, intersect
$\partial \Omega$ at point $A$ and $B$. Here points $C$ and $D$ are chosen to make $ABCD$ be a rectangle.
See
Figure~\ref{positive-a(t)}. By  formula~(\ref{eq:Omega(1)-distance}),
$$\|AD\|=\|xW\|>(G(\Omega)+1)r^2,$$
 then
by formula~(\ref{eq:Remark-equation}) in
Remark~\ref{remark1}, we have that $$\|xA\|=\|DW\|>r.$$
 The distance
between point $x$ and a point $y\in\partial \Omega$ is a continuous
function of $y$, which is due to the smooth
boundary $\partial \Omega$. Notice that $\|xW\|<r$ and
$\|xA\|>r$, so  there exists a
point $E$ in the segment of $\partial \Omega$ from $W$ to $A$ (in an
anticlockwise order), such that the distance between $x$ and $E$ is
just $r$, i.e., $\|xE\|=r$. Similarly, we have another point $F\in \partial \Omega$
between $W$ and $B$ such that $\|xF\|=r$. That is to say, the circle
centred at $x$ with radius $r$ intersects $\partial\Omega$ at points
$E$ and $F$. So the distance from $x$ to  chord $EF$, i.e.,\
$t(x)$, is positive. Thus, $a(t(x))\geq \frac12\pi r^2$, and we
therefore have a lower bound for
$|B(x,r)\cap\Omega|$:
$$|B(x,r)\cap\Omega|\geq a(t(x))\geq \frac12\pi r^2.$$

 \par Now we give an upper bound for $|B(x,r)\cap \Omega|$.
Let line $GH$ be parallel to line $EF$ and be tangent to $\partial \Omega$. In particular,
two points $G$ and $H$ are chosen to make $EFGH$ be a rectangle.
See Figure~\ref{fig:UpperBound-B(x,r)Omega}. The area of this rectangle
$S_{EFGH}\leq\Theta(\|EF\|^3)$, according to
Lemma~\ref{lemma:RectangleArea}. Notice that $\|EF\|$ is less than
$2r$, the diameter of the circle, so we have
 \begin{eqnarray*}
     &&|B(x,r)\cap \Omega|\\
     &\leq& a(t(x))+S_{EFGH}\\
     &\leq & a(t(x))+\Theta(\|EF\|^3)\\
     &\leq& a(t(x))+\Theta(r^3).
\end{eqnarray*}
With $a(t(x))\geq \frac {\pi r^2}2$ that has been derived above, we obtain
\begin{eqnarray*}
     &&|B(x,r)\cap \Omega|\\
     &\leq& a(t(x))\left(1+\frac{\Theta(r^3)}{a(t(x))}\right)\\
     &\leq& a(t(x))(1+o(1)).
\end{eqnarray*}
According to these  lower and upper bounds for $|B(x,r)\cap \Omega|$, we have the
following estimates:
\begin{equation}\label{eq:local-1}
\begin{split}
     &(n|B(x,r)\cap \Omega|)^ke^{-n|B(x,r)\cap \Omega|}\\
     \leq &(1+o(1))^k(na(t(x)))^ke^{-na(t(x))},
\end{split}
\end{equation}
and
\begin{equation}\label{eq:local-2}
\begin{split}
     & (n|B(x,r)\cap \Omega|)^ke^{-n|B(x,r)\cap \Omega|}\\
     \geq &(na(t(x)))^ke^{-n(a(t(x))+\Theta(r^3))}\\
    \geq& e^{-n \Theta(r^3)}(na(t(x)))^ke^{-na(t(x))}.
\end{split}
\end{equation}
\par Consider the integration on $\Omega(1,2)$. Noticing that
$|\Omega(1,2)|\leq lr$ and $\frac12 \pi r^2\leq a(t(x))\leq \pi r^2$, and by formula~(\ref{eq:local-1}), we have
\begin{eqnarray*}
     n\int_{\Omega(1,2)} \psi^k_{n,r}(x)dx&=&\frac n{k!}\int_{\Omega(1,2)}\left(n|B(x,r_n)\cap\Omega|\right)^k e^{-n|B(x,r_n)\cap\Omega|}dx\\
      &\leq&\frac n{k!}\left[(1+o(1))^k(na(t(x)))^ke^{-na(t(x))}\right]|\Omega(1,2)|\\
     &\leq &\frac n{k!}(n\pi r^2)^k e^{-\frac12n\pi r^2}|\Omega(1,2)|\\
     &\leq&\Theta(1)\frac1re^{-\frac12nr^2}\left[(n\pi r^2)^{k+1} e^{-\frac{\pi-1}{2}n\pi r^2}\right]\\
     &=&o(1).
\end{eqnarray*}
That is, we have proved a proposition:
\begin{proposition}\label{pro:Omega(1,2)}The integration of $ \psi^k_{n,r}(x)$ defined by (\ref{eq:psi-function}) on $\Omega(1,2)$ satisfies that
\begin{equation*}
n\int_{\Omega(1,2)} \psi^k_{n,r}(x)dx=o(1).
\end{equation*}
\end{proposition}

\begin{figure}[htbp]
 \begin{center}
       \scalebox{0.55}[0.6]{
           \includegraphics{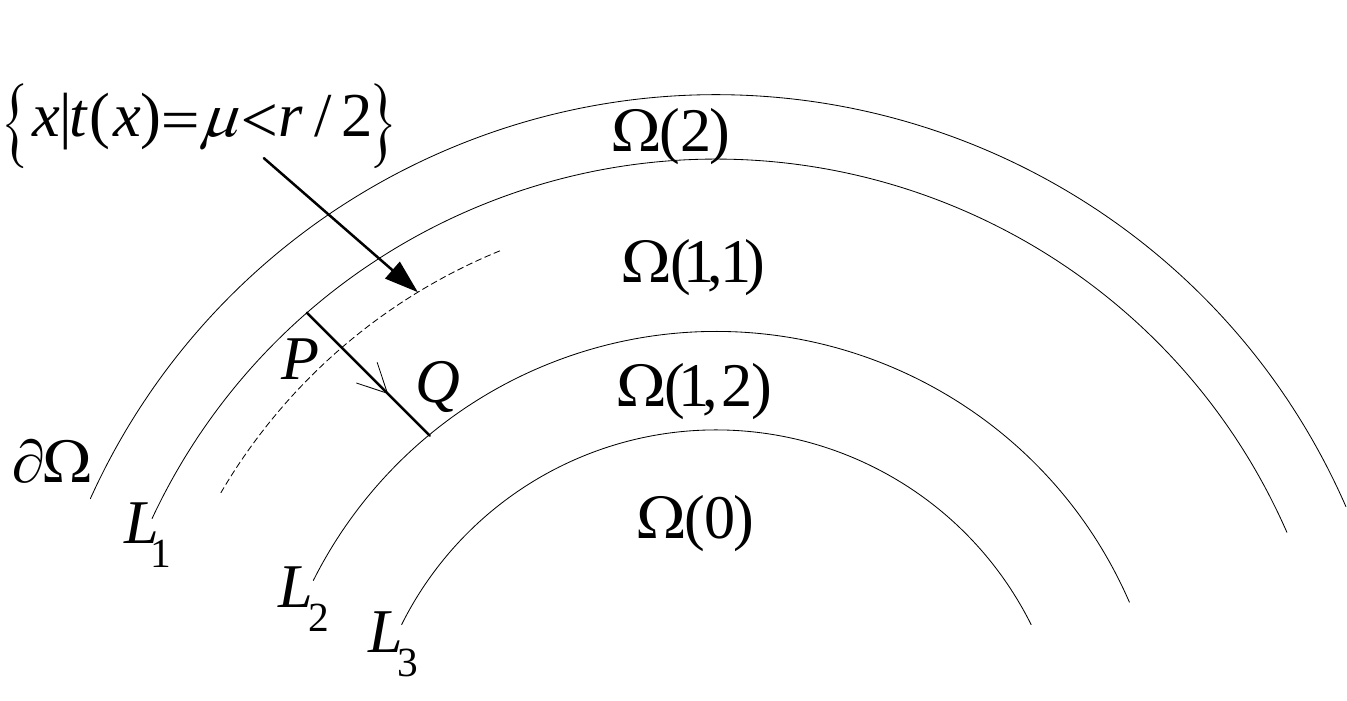}       }
       \caption{$\Omega$ is divided into four parts}\label{fig:Omega(1)}
  \end{center}
  \end{figure}

In the following, we will determine $n\int_{\Omega(1,1)} \psi^k_{n,r}(x)dx$.
By the
estimates (\ref{eq:local-1}) and (\ref{eq:local-2}), we have
\begin{eqnarray*}
     &&\frac n{k!}\int_{\Omega(1,1)}\left(n|B(x,r_n)\cap\Omega|\right)^k e^{-n|B(x,r_n)\cap\Omega|}dx\\
     &\sim &\frac
     n{k!}\int_{\Omega(1,1)}(na(t(x)))^ke^{-na(t(x))}dx.
\end{eqnarray*}

We define
\begin{equation*}\label{eq:L1}
L_1=\{x\in\Omega\mid \mathrm{dist}(x,\partial \Omega)=(G(\Omega)+1)r^2\},
\end{equation*}
\begin{equation*}\label{eq:L2}
L_2=\{x\in\Omega(1)\mid t(x)=\frac{r}{2}\},
\end{equation*}
\begin{equation*}\label{eq:L3}
L_3=\{x\in\Omega\mid \mathrm{dist}(x,\partial \Omega)=r\}.
\end{equation*}
Clearly, the subregion $\Omega(1,1)$ is the strip with the boundaries
 $L_1$ and $L_2$, which is illustrated by
Figure~\ref{fig:Omega(1)}. If $x\in L_1$, then $t(x)\geq 0$ by a similar argument as in the previous
step. Clearly, $t(x)\leq (G(\Omega)+1)r^2$.
 Therefore, $0\leq t(x)\leq (G(\Omega)+1)r^2$ for  any $x\in L_1$. As $r$
 tends to zero, both the curves $L_1$ and $L_2$ approximate the
 boundary $\partial \Omega$. So there exists $\epsilon_r$ such
 that  the length of $L_1$ and $L_2$,
 denoted by $l_1$ and $l_2$ respectively, satisfies:
$$
      l-\epsilon_r\leq l_1,l_2\leq l,
$$
where $\epsilon_r\rightarrow 0$ as $r\rightarrow 0$. In addition, it
is clear that $t(x)$ is increasing along the directed line segment
started from a point in $L_1$ to a point in $L_2$. See the line
segment $PQ$ in Figure~\ref{fig:Omega(1)}.

The integration on the strip $\Omega(1,1)$  can be bounded as
follows:
\begin{equation}\label{eq:Bounds-Integration-Omega(1,1)}
\begin{split}
&(l-\epsilon_r)\frac n{k!}\int_{(G+1)r^2}^{\frac r2}(na(t))^ke^{-na(t)}dt\\
&\leq \frac n{k!}
\int_{\Omega(1,1)}(na(t(x)))^ke^{-na(t(x))}dx\\
&\leq l\frac n{k!}\int_0^{\frac r2}(na(t))^ke^{-na(t)}dt.
\end{split}
\end{equation}

According to Lemma~\ref{lemma:a(t)}, we obtain that
\begin{eqnarray*}
  l\frac n{k!}\int_0^{\frac r2}(na(t))^ke^{-na(t)} dt=\frac{l\sqrt{\pi}}{2^{k+1}k!}e^{-\frac{\xi}2}.
\end{eqnarray*}
Therefore,
\begin{equation*}
\begin{split}
      0&\leq\epsilon_r\int_{(G+1)r^2}^{\frac{r}{2}}\frac
      n{k!}(na(t))^ke^{-na(t)}dt\\
      &\leq \epsilon_r\int_{0}^{\frac{r}{2}}\frac
      n{k!}(na(t))^ke^{-na(t)}dt=o(1).
\end{split}
\end{equation*}
Furthermore, notice that $\pi r^2/2\leq a(t)\leq (\pi+\delta)r^2/2$
for any $t\in [0,(G(\Omega)+1)r^2]$, where $\delta>0$ is a small real number, we
have that
$$
      l\int_0^{(G+1)r^2}\frac n{k!}(na(t))^ke^{-na(t)}dt=o(1).
$$
So,
\begin{equation*}
\begin{split}
&(l-\epsilon_r)\frac n{k!}\int_{(G+1)r^2}^{\frac r2}(na(t))^ke^{-na(t)}dt\\
=&\frac
{n}{k!}\left(l\int_0^{\frac{r}{2}}-l\int_0^{(G+1)r^2}-\epsilon_r\int_{(G+1)r^2}^{\frac
{r}{2}}\right)(na(t))^ke^{-na(t)}dt\\
=&l\frac {n}{k!}\int_0^{\frac{r}{2}}(na(t))^ke^{-na(t)}dt+o(1).
\end{split}
\end{equation*}
Then by formula~(\ref{eq:Bounds-Integration-Omega(1,1)}),
\begin{eqnarray*}
     &&l\frac {n}{k!}\int_0^{\frac{r}{2}}(na(t))^ke^{-na(t)}dt+o(1)\\
     &=&(l-\epsilon_r)\frac n{k!}\int_{(G+1)r^2}^{\frac r2}(na(t))^ke^{-na(t)}dt\\
     &\leq&\frac n{k!}\int_{\Omega(1,1)}(na(t(x)))^ke^{-na(t(x))}dx\\
     &\leq& l\frac n{k!}\int_0^{\frac r2}(na(t))^ke^{-na(t)}dt,
\end{eqnarray*}
which implies that
\begin{eqnarray*}
&&\frac n{k!}\int_{\Omega(1,1)}(na(t(x)))^ke^{-na(t(x))}dx\\
   &\sim& l\frac n{k!}\int_0^{\frac r2}(na(t))^ke^{-na(t)} dt.
\end{eqnarray*}
So, by Lemma~\ref{lemma:a(t)}, we have
\begin{eqnarray*}
     &&\frac n{k!}\int_{\Omega(1,1)}\left(n|B(x,r_n)\cap\Omega|\right)^k e^{-n|B(x,r_n)\cap\Omega|}dx\\
     &\sim &\frac{n}{k!}\int_{\Omega(1,1)}(na(t(x)))^ke^{-na(t(x))}dx\\
     &\sim& l\frac {n}{k!}\int_0^{\frac r2}(na(t))^ke^{-na(t)}dt\\
     &=&\frac{l\sqrt{\pi}}{2^{k+1}k!}e^{-\frac{\xi}2}.
\end{eqnarray*}
Therefore, we have proved the following conclusion:
\begin{proposition}\label{pro:Omega(1,1)} The integration of $ \psi^k_{n,r}(x)$ defined by (\ref{eq:psi-function}) on ${\Omega(1,1)}$ satisfies that
\begin{equation*}
     n\int_{\Omega(1,1)} \psi^k_{n,r}(x)dx=\frac{l\sqrt{\pi}}{2^{k+1}k!}e^{-\frac{\xi}2}.
\end{equation*}
\end{proposition}

Relying on the above four propositions, and noticing that
$\xi$ satisfies
   $$ \left\{\begin{array}{cc}
      e^{-\xi}+\frac{l\sqrt{\pi}}{2^{k+1}k!}e^{\frac{-\xi}{2}}=e^{-c}, & k=1, \\
      \frac{l\sqrt{\pi}}{2^{k+1}k!}e^{\frac{-\xi}{2}}=e^{-c}, & k>1, \\
    \end{array}
        \right.
$$

we finally complete the proof  by the following calculation:
\begin{eqnarray*}
   &&\frac n{k!}\int_{\Omega}\left(n|B(x,r_n)\cap\Omega|\right)^k
   e^{-n|B(x,r_n)\cap\Omega|}dx\\
   &=&\left\{\int_{\Omega(0)}+\int_{\Omega(2)}+\int_{\Omega(1,2)}+\int_{\Omega(1,1)}\right\}
    n\psi^k_{n,r}(x)dx\\
   &\sim& e^{-c}.
\end{eqnarray*}

\par When $k=0$时, notice that $n\int_0^{\frac r2}\frac{(na(t))^ke^{-na(t)}}{k!}dt=o(1)$,
 repeat the above process of the  proof, we know  the conclusion also holds for $k=0$.
 The proof of the first part of Theorem~\ref{thm:combined-Main} in the case
 of smooth $\partial \Omega$ is now completed.

\subsection{Case of continuous $\partial \Omega$}

\begin{figure}[htbp]
 \begin{center}
       \scalebox{0.7}[0.7]{
           \includegraphics{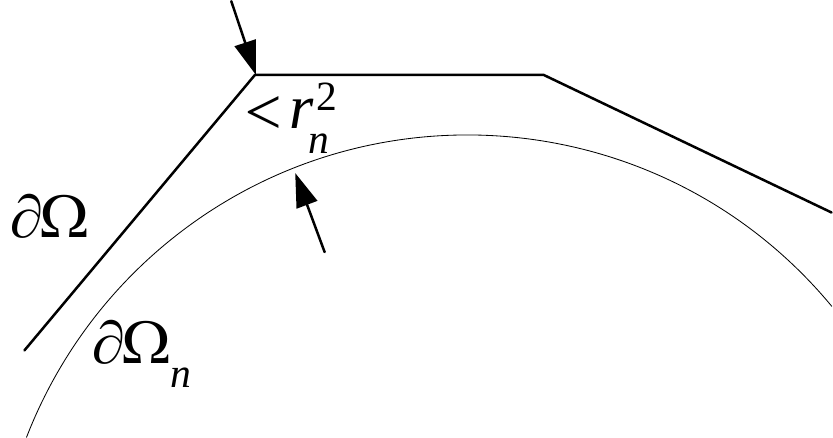}       }
       \caption{Width of the Gap between $\partial \Omega$ and $\partial \Omega_n$ is less than $r_n^2$}\label{fig:Gap}
  \end{center}
  \end{figure}

For a general unit-area convex region $\Omega$ with a continuous
rather than smooth boundary $\partial \Omega$, we may use a family
of convex regions $\{\Omega_n\}_{n=1}^{\infty}\subset \Omega$ to approximate
$\Omega$, where the boundary $\partial \Omega_n$ of each $\Omega_n$
is smooth. We set the width of the gap between $\partial \Omega$ and
$\partial \Omega_n$ to be less than $r^2_n$, see
Figure~\ref{fig:Gap}. That is, $\sup_{x\in \Omega_n}\mathrm{dist}(x,\partial
\Omega)<r^2_n$ for any $n$. Clearly, we have the following
estimation for $S(\Omega_n)$, the area of $\Omega_n$:
$$
    1-lr_n^2\leq S(\Omega_n)\leq 1.
$$

 Because $\partial
\Omega_n$ is smooth, follow the  method presented in the previous
subsection, we can also similarly obtain that
$$
n\int_{\Omega_n}\psi^k_{n,r}(x)dx\sim e^{-c}.
$$
Since the area of $\Omega\setminus \Omega_n$ is no more than $lr_n^2$,
then by the proof of Proposition~\ref{pro:Omega(2)}, we have that
$$
n\int_{\Omega\setminus \Omega_n}\psi^k_{n,r}(x)dx=o(1).
$$
Therefore, we have
$$
n\int_{\Omega}\psi^k_{n,r}(x)dx=n\left(\int_{\Omega_n}+\int_{\Omega\setminus
\Omega_n}\right)\psi^k_{n,r}(x)dx\sim e^{-c}.
$$
This finally  completes the proof of the first part of
Theorem~\ref{thm:combined-Main}.

\section{Proof of Theorem~\ref{thm:combined-Main}: part two}\label{sec:proof-part2}
\label{sec:proof-part2}
{ This section will present a proof for  the second part of Theorem~\ref{thm:combined-Main}. That is, for
$r_n$ given by~(\ref{eq:Theorem-formula-1}) and~(\ref{eq:Theorem-formula-2}) (see Theorem~\ref{thm:combined-Main}), which satisfies
     $$
        \frac n{k!}\int_{\Omega}\left(n|B(x,r_n)\cap\Omega|\right)^k e^{-n|B(x,r_n)\cap\Omega|}dx\sim e^{-c}
     $$
 as we have proved in the previous section, then the probabilities of the two events
$\rho(\chi_n;\delta\geq k+1)\leq r_n$ and $\rho(\chi_n;\kappa\geq
k+1)\leq r_n$ both converge to $\exp\left(-e^{-c}\right)$  as $n\rightarrow\infty$.

Clearly, it is sufficient to prove the following two conclusions:
\begin{conclusion}   Under the assumptions of Theorem~\ref{thm:combined-Main},
\begin{equation}\label{eq:conclusion-1}
\lim_{n\rightarrow\infty}\Pr\left\{\rho(\chi_n;\delta\geq k+1)\leq r_n\right\} =e^{-e^{-c}}.
\end{equation}
\end{conclusion}

\begin{conclusion} Under the assumptions of Theorem~\ref{thm:combined-Main},
\begin{equation}\label{eq:conclusion-2}
\lim_{n\rightarrow\infty}\Pr\left\{\rho(\chi_n;\delta\geq k+1)=\rho(\chi_n;\kappa\geq
k+1)\right\}=1.
\end{equation}
\end{conclusion}

We will follow Penrose's approach to prove these two conclusions. This section mainly presents a proof for Conclusion~1, while we
sketch the proof of Conclusion~2 in Appendix~C. The approach  is instead to prove  a Poissonized version of Conclusion~1:
\begin{equation}\label{eq:concusion-1-Poisson}
\lim_{n\rightarrow\infty}\Pr\left\{\rho(\mathcal{P}_n;\delta\geq k+1)\leq r_n\right\} =\exp\left(-e^{-c}\right),
\end{equation}
where $\mathcal{P}_n$ means that the $n$ points distributed over unit-area convex region $\Omega$ are Poisson point process, which
has been introduced in Section~\ref{sec-related-mainresults}.  Because $\chi_n$ can be well approximated by $\mathcal{P}_n$,
it is naturally  to see the original (\ref{eq:conclusion-1}) hold by a de-Poissonized technique.
The de-Poissonized technique is standard and thus omitted here, please see~\cite{Penrose-k-connectivity} for details. In the following, we only demonstrate
the Poissonized version (\ref{eq:concusion-1-Poisson}) holds for a general convex region $\Omega$.

 For given $n$, $x,y\in \Omega$, let
  $$
   v_x=|B(x,r)\cap \Omega |, v_y=|B(y,r)\cap \Omega |, $$
   $$v_{x,y}=|B(x,r)\cap B(y,r)\cap \Omega|,$$
   $$v_{x\backslash y}=v_x-v_{x,y},v_{y\backslash x}=v_y-v_{x,y}.
  $$
Define $I_i=I_i(n)(i=1,2,3)$ as follows:
$$
    I_1=n^2\int_{\Omega} dx\int_{\Omega\cap B(x,3r)}dy \psi^k_{n,r}(y)\psi^{k}_{n,r}(x),
$$
$$
   I_2=n^2\int_{\Omega} dx\int_{\Omega\cap B(x, r)}dy \Pr[Z_1+Z_2=Z_1+Z_3=k-1],
$$
$$
   I_3=n^2\int_{\Omega} dx\int_{\Omega\cap B(x,3r)\backslash B(x,r)}dy \Pr[Z_1+Z_2=Z_1+Z_3=k],
$$
where $Z_1,Z_2,Z_3$ are independent Poisson variables with means
$nv_{x,y},nv_{x\backslash y},nv_{y\backslash x}$ respectively. As Penrose has pointed out in~\cite{Penrose-k-connectivity},
by an argument similar to that of Section~7 of~\cite{Penrose97:longest-edge}, it suffices to prove that
$I_1,I_2,I_3\rightarrow 0$ as $n\rightarrow \infty$.
}

%

 However, some critical estimations  in Penrose's proof for $I_1,I_2,I_3\rightarrow 0 (n\rightarrow \infty)$
 are only  valid for a regular region square (see~\cite{Penrose-RGG-book,Penrose-k-connectivity}).
Our work, i.e., Theorem~\ref{thm:combined-Main} or Conclusion~1,   extends these results to the case of a general domain, i.e., a general convex region.
For this purpose, we will first establish a proposition, an estimation in the case of  general convex region $\Omega$.


\begin{proposition}\label{proposition1} Under the assumptions of Theorem~\ref{thm:combined-Main}, $Z_3$ is a Poisson variable with mean $nv_{y\backslash x}$ defined above, then

  \begin{equation*}
\frac{1}{n\pi r^2}\int_{\Omega}n\psi^k_{n,r}(x)dx\int_{\Omega\cap
B(x,r)}n\Pr(Z_3=k-1)dy=o(1).
\end{equation*}
\end{proposition}
\begin{proof} The proof is presented in Appendix~B.

\end{proof}

For the remainder proof of Conclusion~1, as we mentioned, it is sufficient to prove $I_1,I_2,I_3\rightarrow 0$, as
$n\rightarrow 0$. This  will be completed in the following three claims.

\begin{claim}
$I_1\rightarrow 0$ as $n\rightarrow \infty$.
\end{claim}
\begin{proof}
By Lemma~\ref{lemma:LowerBound-B(x,r)}, there exists a constant $C>0$  such that $C\pi r^2\leq |B(x,r)\cap\Omega|\leq \pi r^2$, so
$$
          \frac{(Cn\pi r^2)^k\exp(-n\pi r^2)}{k!}\leq \psi^k_{n,r}\leq \frac{(n\pi r^2)^k\exp(-Cn\pi r^2)}{k!},
$$
and thus,
{
 \begin{eqnarray*}
&&I_1\\
&=&n^2\int_{\Omega} dx\int_{{\Omega}\cap B(x,3r)}dy \psi^k_{n,r}(y)\psi^{k}_{n,r}(x)\\
    &=& \int_{\Omega} n\psi^{k}_{n,r}(x)dx \int_{B(x,3r)\cap \Omega}n\psi^{k}_{n,r}(y)dy\\
    &\leq& \int_{\Omega} n\psi^{k}_{n,r}(x)dx \int_{B(x,3r)\cap \Omega}\frac{1}{k!} n(n\pi r^2)^k\exp(-Cn\pi r^2)\\
    &\leq & \frac{n}{k!}|B(x,3r)|(n\pi r^2)^{k}\exp(-Cn\pi r^2)\int_{\Omega} n\psi^{k}_{n,r}(x)dx\\
    &=& \frac{9}{k!}(n\pi r^2)^{k+1}\exp(-Cn\pi r^2)\int_{\Omega} n\psi^{k}_{n,r}(x)dx\\
    &\sim& e^{-c}\frac{9}{k!}(n\pi r^2)^{k+1}\exp(-Cn\pi r^2)\\
    &\longrightarrow & 0,n\longrightarrow\infty.
\end{eqnarray*}
The last one holds due to   Remark~\ref{rem:remark3}, i.e.,  $r_n$ given by (\ref{eq:Theorem-formula-1}) and (\ref{eq:Theorem-formula-2}) in Theorem~\ref{thm:combined-Main} satisfying
$$
     (n\pi r^2)^{k+1}\exp(-Cn\pi r^2)\sim \frac{(\log n+o(\log n))^{k+1}}{e^{C(\log n+o(\log n))}}=o(1).
$$
}
\end{proof}

\begin{claim}
$I_2\rightarrow 0$, as $n\rightarrow \infty$.
\end{claim}
\begin{proof}
{  It is obvious that
$Z_1+Z_2$ and $Z_3$ are Possison variables with means $nv_x$ and
$nv_{y\backslash x}$ respectively.
Notice that
\begin{eqnarray*}
&&\Pr[Z_1+Z_3=k-1| Z_1+Z_2=k-1]\\
&=& \sum_{j=0}^{k-1}\Pr[Z_1+Z_3=k-1;Z_1=j|Z_1+Z_2=k-1]\\
&\leq &\sum_{j=0}^{k-1}\Pr[Z_3=k-1-j |Z_1+Z_2=k-1]\\
&=&\sum_{j=0}^{k-1}\Pr[Z_3=k-1-j].
\end{eqnarray*}

For any $x\in\Omega$,
 $$\psi^{k}_{n,r}(x)=\frac{n|B(x,r)\cap \Omega|}{k}\psi^{k-1}_{n,r}(x),$$
then by Lemma~\ref{lemma:LowerBound-B(x,r)}, there exists a constant $C>0$ such that
$C\pi r^2\leq |B(x,r)\cap \Omega|$, and thus
$$\psi^{k-1}_{n,r}(x)\leq \frac{k}{Cn\pi r^2}\psi^{k}_{n,r}(x).$$
Therefore, we have
\begin{eqnarray*}
    &&I_2\\
    &=&n^2\int_{\Omega} dx\int_{\Omega\cap B(x,r)}dy \Pr[Z_1+Z_2=Z_1+Z_3=k-1]\\
       &=&n^2\int_{\Omega} dx\int_{\Omega\cap B(x,r)}dy
       \Pr[Z_1+Z_2=k-1]\\
       && \Pr[Z_1+Z_3=k-1|Z_1+Z_2=k-1]\\
       &=& n \int_{\Omega}\Pr[Z_1+Z_2=k-1] dx\\
       &&\int_{\Omega\cap B(x,r)}n\Pr[Z_1+Z_3=k-1|Z_1+Z_2=k-1]dy\\
       &\leq & n\int_{\Omega}\Pr[Z_1+Z_2=k-1] dx\\
       && \left\{\sum_{j=0}^{k-1}\int_{\Omega\cap B(x,r)}n\Pr[Z_3=k-1-j]dy\right\}\\
       &=& \sum_{j=0}^{k-1} \int_{\Omega}n\psi^{k-1}_{n,r}(x) dx \int_{\Omega\cap B(x,r)}n\Pr[Z_3=k-1-j]dy\\
       &\leq& \frac{C}{k} \sum_{j=0}^{k-1}
       \frac{1}{n\pi r^2}\int_{\Omega}n\psi^{k}_{n,r}(x) dx \int_{\Omega\cap B(x,r)}n\Pr[Z_3=k-1-j]dy\\
       &=&o(1).
\end{eqnarray*}
The last equation holds due to Proposition~\ref{proposition1} and the proved conclusion $\int_{\Omega}n\psi^{k}_{n,r}(x) dx \sim
e^{-c}$.
}
\end{proof}

\begin{claim}
$I_3\rightarrow 0$, as $n\rightarrow \infty$.
\end{claim}
\begin{proof}
 The proof is
similar to that of  Claim 2.
\end{proof}

Therefore, by these three claims,  we know the Poissonized version (\ref{eq:concusion-1-Poisson}) holds, which leads to
Conclusion~1 by a de-Poissonized techinique as we have mentioned:
 $$\lim_{n\rightarrow\infty}\Pr\left\{\rho(\chi_n;\delta\geq k+1)\leq r_n\right\} =e^{-e^{-c}}.$$
The proof of Conclusion~2 is presented in Appendix~C.
As long as these two conclusion are proved, the second part of  Theorem~\ref{thm:combined-Main} is consequently proved.
Therefore,  the proof of Theorem~\ref{thm:combined-Main} is completed.


\section{Conclusions and Future Work}
\label{sec-conclusion}

In this article, we have studied a more general unit area convex region on which wireless sensor and ad-hoc networks are distributed. We have investigated two types of critical transmission radii in the networks, i.e.,\ $\rho(\chi_n;\kappa\geq k)$ and $\rho(\chi_n;\delta\geq k)$,  defined in terms of $k-$connectivity and the minimum vertex degree respectively. We have also established the precise asymptotic distributions of these two types of radii, which makes the previous results obtained under the circumstance of squares or disks special cases of this work. More importantly, our results have clearly indicated that only the length of the boundary of regions completely determines the critical transmission radius. Further, by isodiametric inequality, the smallest asymptotic critical radius  is achieved only when the region  is a disk.

\section*{Acknowledgments}
J. Ding acknowledges the financial support by the NSF of China under Grant 61472343.
S. Ma is supported in part by NSF of China under Grant 61925203.
X. Zhu is supported by NSF of China under Grant 61972282.

\appendices
 \setcounter{proposition}{4}
 \setcounter{lemma}{0}
  \setcounter{figure}{7}

\section{Proofs of Lemma~1 and~2 }

\begin{lemma}\label{lemma:LowerBound-B(x,r)} Let $\Omega$  be a bounded convex region,
then there exists a positive constant $C$ such that for any
sufficiently small $r$,
$$
        \mathop{\inf}_{x\in \Omega}|B(x,r)\cap \Omega |\geq C\pi
        r^2.
$$
\end{lemma}

\begin{figure}[htbp]
 \begin{center}
       \scalebox{0.7}[0.7]{
           \includegraphics{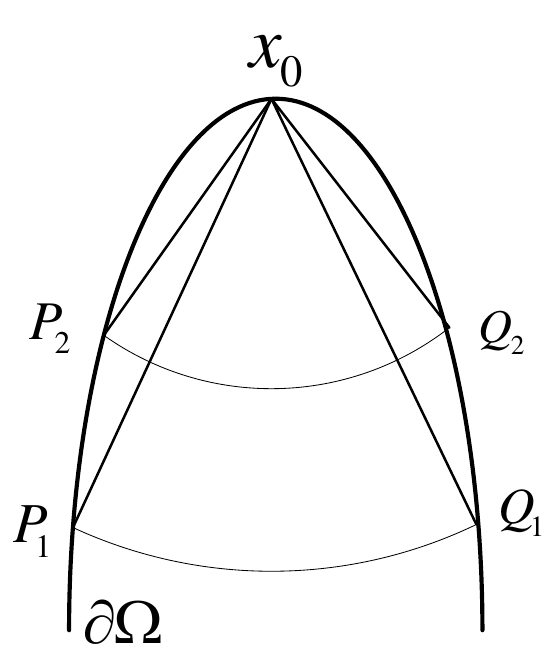} }
       \caption{Sectors on $\overline{\Omega}$}\label{fig:Lemma1}
  \end{center}
  \end{figure}

\begin{proof}
Define
$$
  f(x,r)=\frac{\left|B(x,r)\cap \Omega\right|}{\pi r^2},
$$
then $f(x,r)$ is a continuous function of $x$ and $r$ on $\overline{\Omega}\times (0, 1]$, where $\overline{\Omega}$ is the closure of $\Omega$.
We will show that there exists a constant $C$ such that $f(x,r)>C>0$ on $\overline{\Omega}\times (0, 1]$. Otherwise,
we assume there exists a sequence  $\{(x_n,r_n)\}_{n=1}^{\infty}$ such that
$f(x_n,r_n)\rightarrow 0$ as $n$ tends to infinity.
Because $\overline{\Omega}$ is a bounded and closed  region,  there exists a subsequence of
$\{x_n\}_{n=1}^\infty$, namely $\{x_{n_k}\}_{k=1}^{\infty}$, converging to a limit, denoted by $x_0\in\overline{\Omega}$, as $k$ tends to infinity.
Therefore,
$$
     \lim_{k\rightarrow \infty}f(x_{0},r_{n_k})=\lim_{k\rightarrow \infty}f(x_{n_k},r_{n_k})=0.
$$

(1) If $x_0\in \Omega^{o}$, i.e., $x_0$ is an interior point in $\Omega$, then there exists
a $\delta>0$ such that $B(x_0, \delta)\subset \Omega^{o}\subset \Omega$. So
for any $r<\delta$, we always have $f(x_0,r)=1$, which contradicts to $\lim_{k\rightarrow \infty}f(x_{0},r_{n_k})=0$.

(2) Otherwise, $x_0\in \partial \Omega$. For  small $r'>r''>0$, suppose the circle with the center $x_0$ and radius $r'$ intersects the
boundary $\partial \Omega$ at $P_1$ and $Q_1$ (see Figure~\ref{fig:Lemma1}), then we obtain a
sector $P_1x_0Q_1$, denoted by $S(x_0,r')$. Because $\Omega$ is a convex region, so $S(x_0,r')\subset \overline{\Omega}$ as long as $r'$ is sufficiently small. The angle of sector $S(x_0,r')$ is denoted by $\theta(x_0,r')$.
Similarly, we assume the circle centering $x_0$ with radius $r''$ intersects the
boundary $\partial \Omega$ at $P_2$ and $Q_2$, then we obtain a
sector $P_2x_0Q_2$, denoted by $S(x_0,r'')$. The angle of sector $S(x_0,r'')$ is denoted by $\theta(x_0,r'')$.
We also know $S(x_0,r'')\subset \overline{\Omega}$.  Notice that $r''<r'$, by the convexity of $\overline{\Omega}$, it is easy to
see that $\theta(x_0,r'')\geq \theta(x_0,r')$.
Therefore,  for  any sufficiently small $r<r'$,
\begin{eqnarray*}
|B(x_0,r)\cap\Omega|&\geq& |S(x_0,r)\cap\Omega|\\
&=& |S(x_0,r)|\\
&=&\frac{\theta(x_0,r)}{2\pi}\pi r^2 \\
&\geq& \frac{\theta(x_0,r')}{2\pi}\pi r^2,
\end{eqnarray*}
leading to
$$
  f(x_0,r)=\frac{\left|B(x_0,r)\cap \Omega\right|}{\pi r^2}\geq \frac{\theta(x_0,r')}{2\pi},
$$
which contradicts to $\lim_{k\rightarrow \infty}f(x_{0},r_{n_k})=0$. The proof of this lemma is complete.
\end{proof}

We should point out that if the boundary $\partial \Omega$ is smooth, then we can further obtain that
the constant $C$ can approximate $\frac{1}{2}$ when $r$ tends to zero.

\begin{figure}[ht]
\centering
\includegraphics[width=5cm]{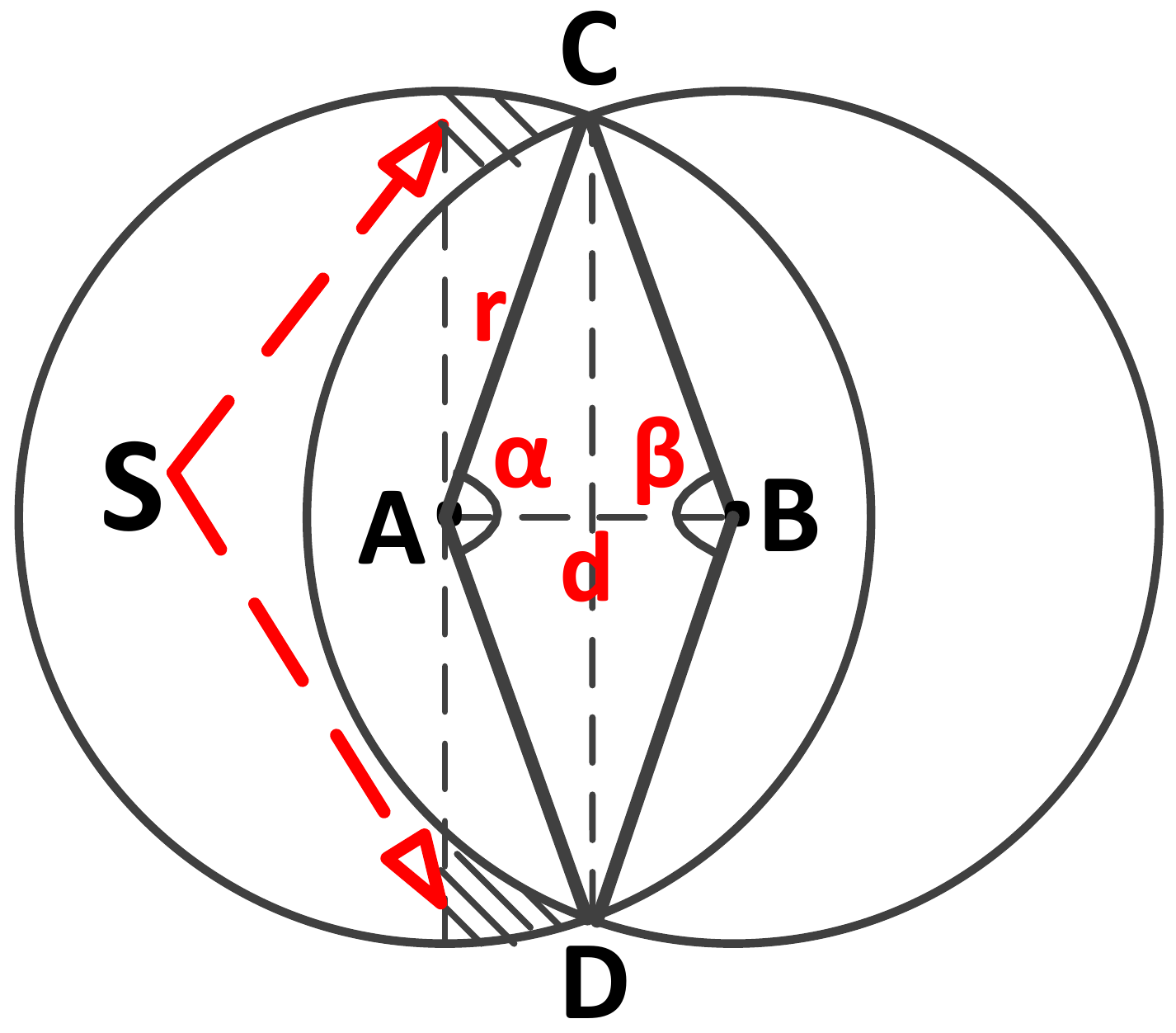}
\caption{Illustration of Lemma~\ref{lem:shadow-low-bound}: Area $S$ }
\label{fig:Figure5-1-Appendix}
\end{figure}

\begin{lemma}\label{lem:shadow-low-bound}
Two disks centered at $A$ and $B$ respectively, have  the same radii $r$, with the distance $\|AB\|=d$.
The shadow area $S$ in
Figure~\ref{fig:Figure5-1-Appendix} is
$$S=r^2 \left[\frac{1}{2}\frac{d}{r} + \frac{19}{48} \left(\frac{d}{r}\right)^3 + \frac{153}{1280}\left(\frac{d}{r}\right)^5
- o\left(\frac{d}{r}\right)^5\right].$$
In particular, if $r\geq d\geq \frac{r}{(nr^2)^{\frac12}}=\frac{1}{\sqrt{n}}$, then
$$nS\geq \frac{1}{\sqrt{\pi}}\left(n\pi r^2\right)^{\frac12}.$$
\end{lemma}

\begin{figure}[ht]
\centering
\includegraphics[width=5cm]{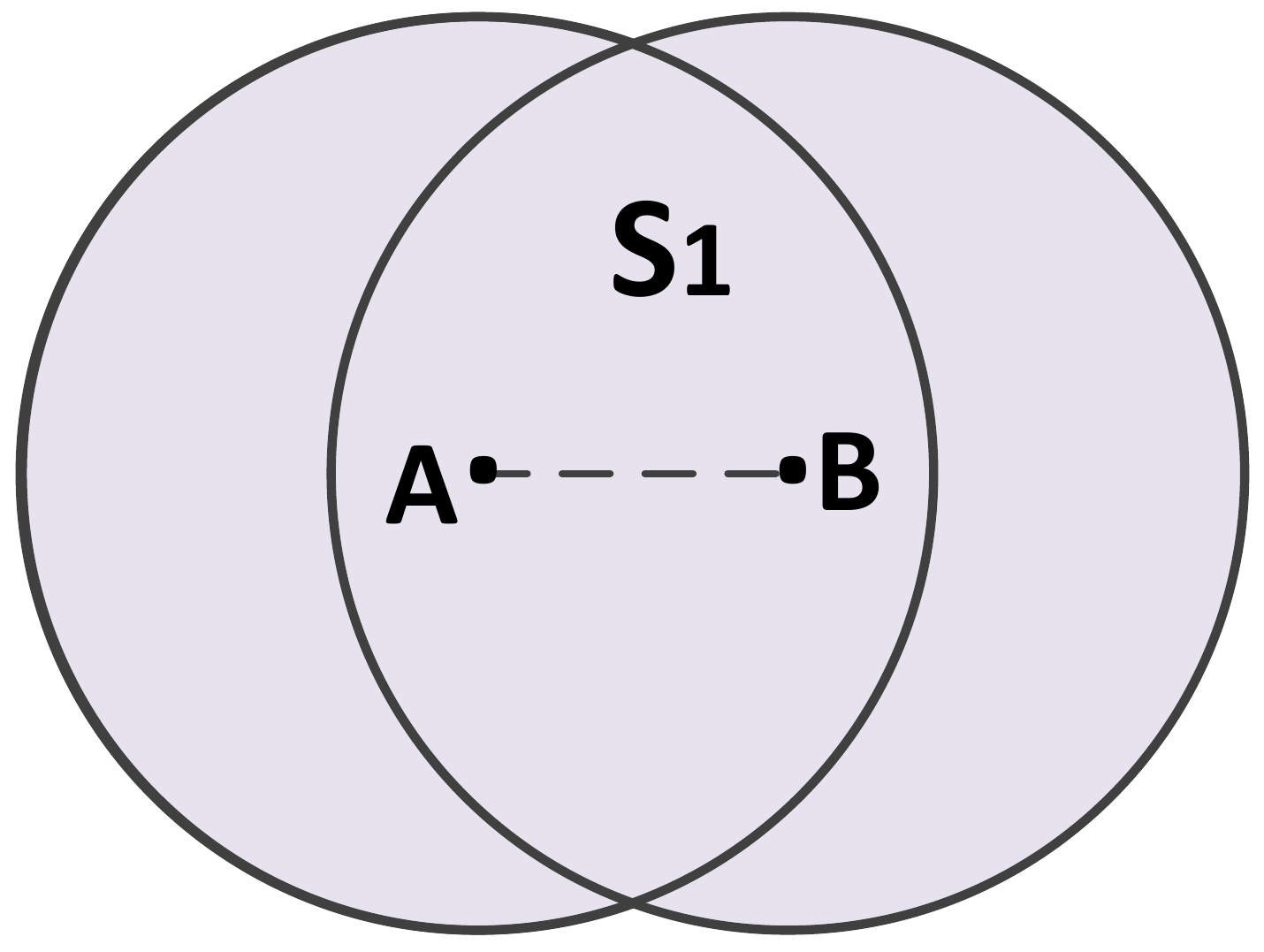}
\caption{Illustration of Lemma~\ref{lem:shadow-low-bound}: Area $S_1$ }
\label{fig:Figure5-2}
\end{figure}

\begin{figure}[ht]
\centering
\includegraphics[width=5cm]{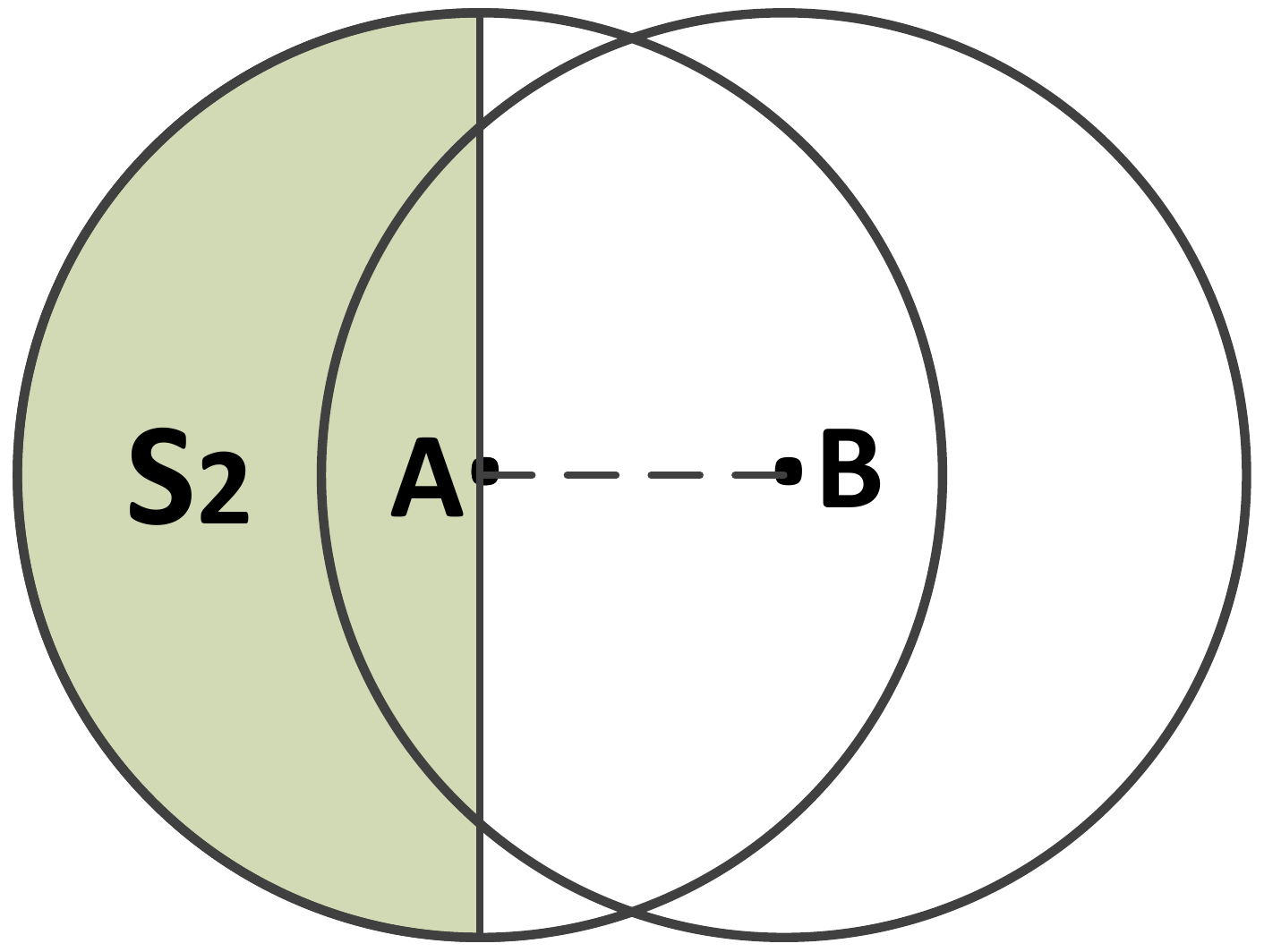}
\caption{Illustration of Lemma~\ref{lem:shadow-low-bound}: Area $S_2$ }
\label{fig:Figure5-3}
\end{figure}

\begin{figure}[ht]
\centering
\includegraphics[width=5cm]{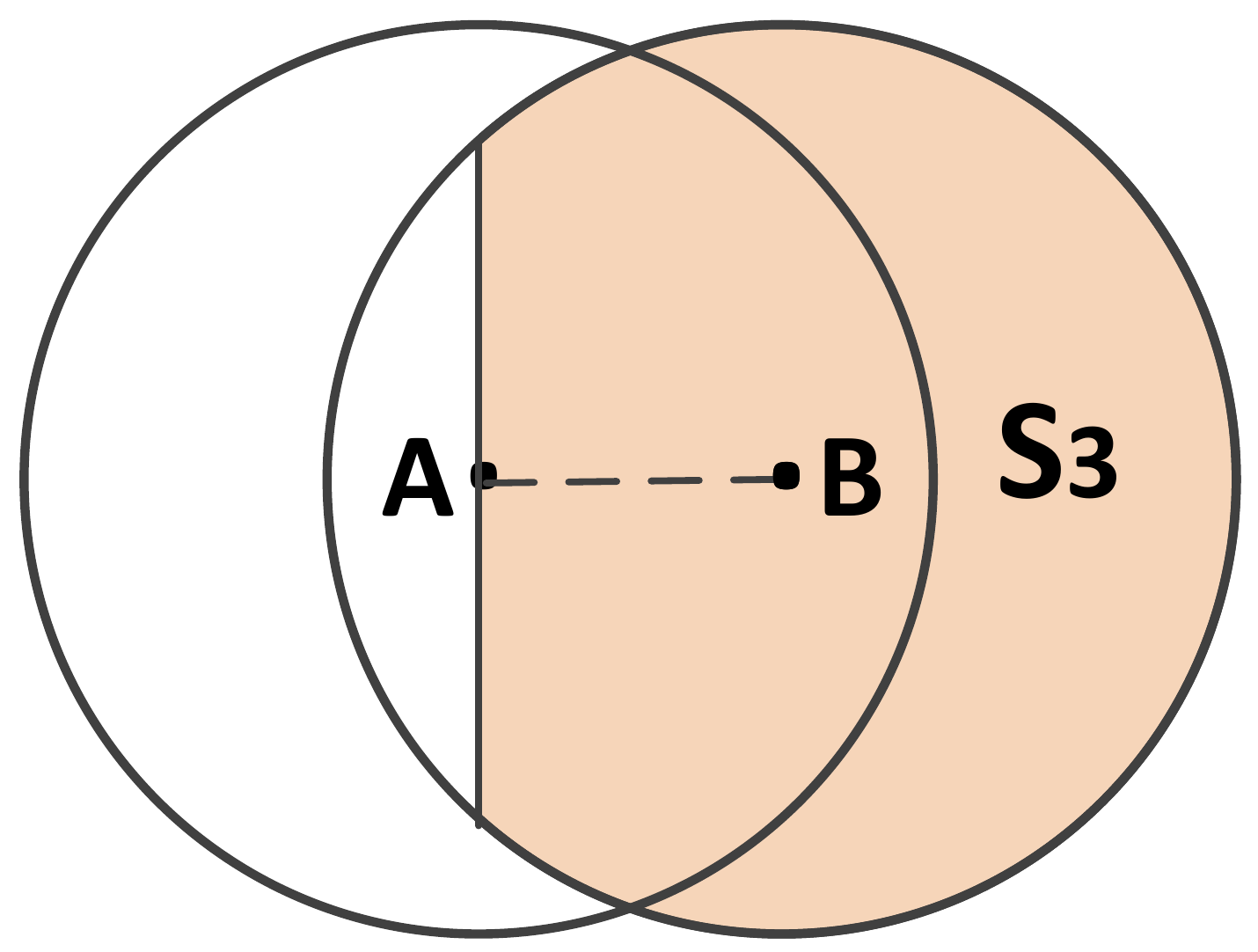}
\caption{Illustration of Lemma~\ref{lem:shadow-low-bound}: Area $S_3$ }
\label{fig:Figure5-4}
\end{figure}

\begin{proof}
It is clear that  quadrilateral $ABCD$ is a diamond, with the angles $\alpha = \beta$. 
Let $S_1$ be the total area of these two overlap disks (see Figure~\ref{fig:Figure5-2}), $S_2$ be the area of gray semi-disk (see Figure~\ref{fig:Figure5-3}), and
$S_3$ be the area of the acid blue region (see Figure~\ref{fig:Figure5-4}). Then, it is easy to see
\begin{align*}
S_1 & =2\pi r^2 - 2\left[\pi r^2 \frac{\alpha}{2\pi} - \frac{d}{2} \sqrt{r^2 - \left(\frac{d}{2}\right)^2}\right],\\
S_2 & =\frac{\pi r^2}{2},\\
S_3 & =\pi r^2 - \left[\pi r^2 \frac{\beta}{2 \pi} - d \sqrt{r^2 - d^2}\right],
\end{align*}
and the shadow area
\begin{align}
S & =S_1 - S_2 - S_3 \nonumber\\
&  = \frac{\pi r^2}{2} + d \sqrt{r^2 -\left(\frac{d}{2}\right)^2} - d \sqrt{r^2 - d^2} + \pi r^2\left(\frac{\beta}{2\pi} - \frac{\alpha}{\pi}\right)\nonumber\\
& = \frac{\pi r^2}{2} + d \sqrt{r^2 -\left(\frac{d}{2}\right)^2} - d \sqrt{r^2 - d^2} + \pi r^2\left(\frac{\beta}{2\pi} - \frac{\beta}{\pi}\right)\nonumber\\
& = \frac{\pi r^2}{2} + d \sqrt{r^2 -\left(\frac{d}{2}\right)^2} - d \sqrt{r^2 - d^2} - \frac{ r^2\beta}{2}.
\end{align}

%
%
%
%

According to Figure~\ref{fig:Figure5-1-Appendix},
\begin{align*}
\cos \left(\frac{\beta}{2}\right) & = \frac{d}{2r},
\end{align*}
or
\begin{align*}
\frac{\beta}{2} & = \arccos \left(\frac{d}{2r}\right).
\end{align*}
Notice that
$
\arccos (x) = \frac{\pi}{2} - x - \frac{x^3}{6} - \frac{3x^5}{40} + o (x^5).
$
So,
\begin{align*}
\frac{\beta}{2} & = \arccos \left(\frac{d}{2r}\right)\\
& = \frac{\pi}{2} - \frac{d}{2r} - \frac{\left(\frac{d}{2r}\right)^3}{6} - \frac{3\left(\frac{d}{2r}\right)^5}{40} + o \left(\frac{d}{2r}\right)^5,
\end{align*}
which leads to
\begin{align*}
 \frac{r^2 \beta}{2}= \frac{\pi}{2} r^2 - r^2 \left[\frac{1}{2}\frac{d}{r} + \frac{1}{48}  \left(\frac{d}{r}\right)^3 + \frac{3}{1280}  \left(\frac{d}{r}\right)^5 - o\left(\frac{d}{r}\right)^5\right].
\end{align*}
It is clearly,
$$
    \sqrt{1 - \left(\frac{d}{r}\right)^2} =1-\frac{1}{2} \left(\frac{d}{r}\right)^2-\frac{1}{8}\left(\frac{d}{r}\right)^4+o\left(\left(\frac{d}{r}\right)^5\right),
$$
$$
    \sqrt{1 - \left(\frac{d}{2r}\right)^2} =1-\frac{1}{2}\left(\frac{d}{2r}\right)^2-\frac{1}{8}\left(\frac{d}{2r}\right)^4+o\left(\left(\frac{d}{r}\right)^5\right).
$$
So the shadow area $S$ can be written as
\begin{align*}
S &= \frac{\pi r^2}{2} + dr \sqrt{1 -\left(\frac{d}{2r}\right)^2} - dr \sqrt{1-\left(\frac{d}{r}\right)^2} - \frac{ r^2\beta}{2}     \\
& =  dr \left[\frac{3}{8}\left(\frac{d}{r}\right)^2+\frac{15}{128}\left(\frac{d}{r}\right)^4\right]\\
 &+  r^2 \left[\frac{1}{2}\frac{d}{r} + \frac{1}{48} \left(\frac{d}{r}\right)^3 + \frac{3}{1280} \left(\frac{d}{r}\right)^5
  - o\left(\frac{d}{r}\right)^5\right]\\
 &= r^2 \left[\frac{1}{2}\frac{d}{r} + \frac{19}{48} \left(\frac{d}{r}\right)^3 + \frac{153}{1280} \left(\frac{d}{r}\right)^5
 - o\left(\frac{d}{r}\right)^5\right].
\end{align*}
Clearly, if $r\geq d\geq \frac{r}{(nr^2)^{\frac12}}=\frac{1}{\sqrt{n}}$, we have
\begin{equation*}
nS  > \frac{1}{2}nr^2\frac{d}{r}\geq\frac{1}{\sqrt{\pi}}\left(n\pi r^2\right)^{\frac12}.
\end{equation*}

%
%
%
%
\end{proof}

\section{Proof of Proposition~\ref{proposition1}}
As we have denoted previously,
  $$
   v_y=|B(y,r)\cap \Omega |, v_{x,y}=|B(x,r)\cap B(y,r)\cap \Omega|,$$
  $$ v_{y\backslash x}=v_y-v_{x,y}.
  $$
$Z_3$ is a Poisson variable with mean $nv_{y\backslash x}$, and
$$
       \psi^k_{n,r}(x)= \frac{\left(n|B(x,r)\cap\Omega|\right)^k
       e^{-n|B(x,r)\cap\Omega|}}{k!}.
$$

Now we prove the following
\begin{proposition} Let $\Omega$ be an unit-area convex region and $r$ be given by~(\ref{eq:Theorem-formula-1}) and ~(\ref{eq:Theorem-formula-2})   in Theorem~\ref{thm:combined-Main}, then
  \begin{equation*}
\frac{1}{n\pi r^2}\int_{\Omega}n\psi^k_{n,r}(x)dx\int_{\Omega\cap
B(x,r)}nP(Z_3=k-1)dy=o(1).
\end{equation*}
\end{proposition}

\begin{proof} Let $d_0=\frac{r}{\left(n\pi r^2 \right)^\frac{1}{2}}$, and
            $\forall x\in \Omega$, let
 $$
       \Gamma_1(x)=\left\{y\in B(x,r)\cap \Omega: \mathrm{dist}(y,x)\leq d_0=\frac{r}{\left(
                    n\pi r^2
            \right)^\frac{1}{2}}\right\},$$
\begin{equation}
\begin{split}
 &\Gamma_2(x)=\\
 &\left\{ y\in B(x,r)\cap \Omega: \mathrm{dist}(y,x)\geq d_0, \mathrm{dist}(y,\partial \Omega)> (G(\Omega)+1)r^2\right\},
\end{split}
\end{equation}
$$
    \Gamma_3(x)=\left\{y\in B(x,r)\cap \Omega: \mathrm{dist}(y,\partial \Omega)\leq (G(\Omega)+1)r^2\right\}.
$$
Here $\mathrm{dist}(y,x)$ denotes the distance between points $y$ and $x$, i.e., $\|yx\|$, and $\mathrm{dist}(y,\partial \Omega)=\inf_{x\in\partial \Omega}\mathrm{dist}(y,x)$, is the
distance between point $y$ and curve $\partial \Omega$.
{ Constant $G(\Omega)$ is an uniform upper bound of all
second-order derivatives of the function of $\partial \Omega$ in local coordinates, as discussed in Remark~\ref{remark1}. }
Clearly, as Figure~\ref{fig:Ball-in-Gammas} illustrates,
$$
    B(x,r)\cap\Omega\subset\Gamma_1(x)\cup\Gamma_2(x)\cup\Gamma_3(x).
    $$
\begin{figure}[htbp]
 \begin{center}
       \scalebox{0.7}[0.7]{
           \includegraphics{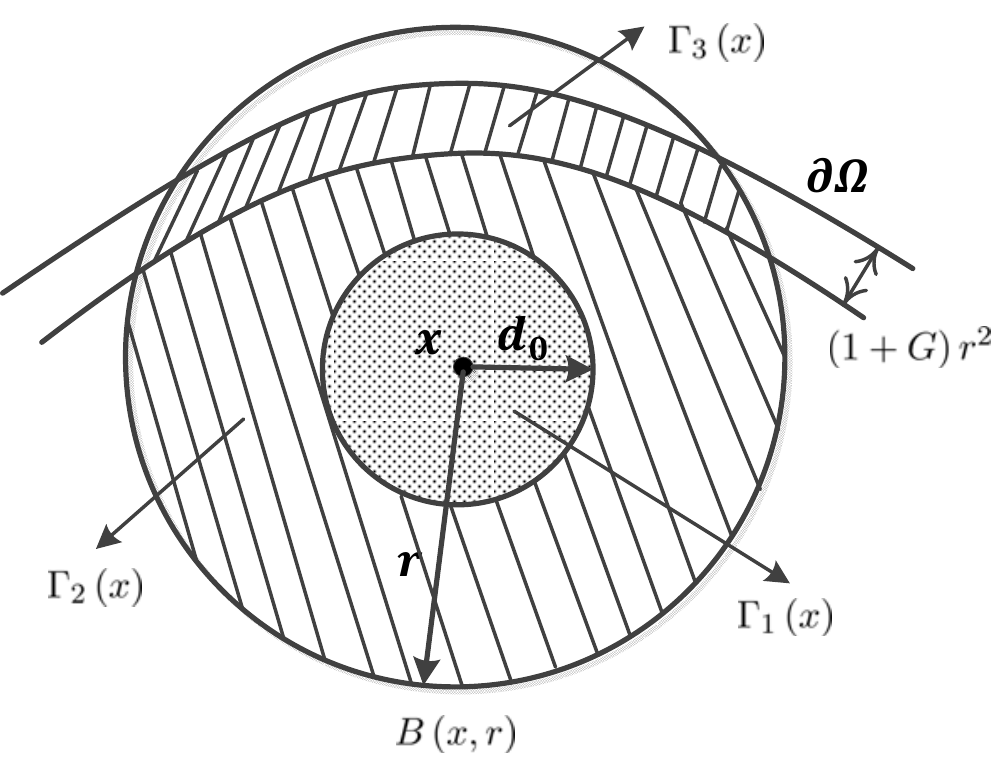}       }
       \caption{Illustration of $B(x,r)\cap\Omega\subset\Gamma_1(x)\cup\Gamma_2(x)\cup\Gamma_3(x)$}\label{fig:Ball-in-Gammas}
  \end{center}
  \end{figure}

{ Notice that $\Gamma_3(x)$ falls in a region with the width less than $(G(\Omega)+1)r^2$ and the length less than $2\pi r$ (the perimeter of $B(x,r)$), therefore
$$    |\Gamma_3(x)|\leq 2\pi(G(\Omega)+1)r^3. $$
}
Let
\begin{equation}\label{eq:A123}
      A_i=\frac{1}{n\pi r^2}\int_{\Omega}n\psi^k_{n,r}(x)dx\int_{\Gamma_i(x)}n\Pr(Z_3=k-1)dy, i=1,2,3.
\end{equation}

{ We will prove $A_i=o(1), i=1,2,3$, in the following three steps. Notice that
we have proved in Section~\ref{sec:proof-part1}: $\int_{\Omega}n\psi^k_{n,r}(x)dx\sim e^{-c}$.}

\par Step 1: to prove $A_1=o(1)$ where $A_1$ is defined by (\ref{eq:A123}).
\begin{eqnarray*}
    A_1&=&\frac{1}{n\pi r^2}\int_{\Omega}n\psi^k_{n,r}(x)dx\int_{\Gamma_1(x)}n\Pr(Z_3=k-1)dy\\
    &\leq&\frac{1}{n\pi r^2}\int_{\Omega}n\psi^k_{n,r}(x)dx\int_{\Gamma_1(x)}ndy\\
     &\leq& \frac{1}{\pi r^2}\int_{\Omega}|\Gamma_1(x)|n\psi^k_{n,r}(x)dx\\
       &\leq& \frac{\pi d_0^2}{\pi r^2}\int_{\Omega}n\psi^k_{n,r}(x)dx\\
       &\sim&\frac{1}{(n\pi r^2)^\frac{1}{2}}e^{-c}\\
       &=&o(1).
\end{eqnarray*}

\par Step 2: to prove $A_2=o(1)$ where $A_2$ is defined by (\ref{eq:A123}).
{ According to the definition of  $\Gamma_2(x)$, for any $y\in \Gamma_2(x)$, the distance from $y$ to $\partial \Omega$ is larger than
$(G(\Omega)+1)r^2$. Then by Remark~2, $\|yA\|>r$ and $\|yB\|>r$, see Figure~\ref{fig:lemma2-Application}. This means that at least more than one half of  $B(y,r)$ falling in $\Omega$. So there exists at lest one joint point of $\partial B(y,r)$ and $\partial B(x,r)$ falls in
$\Omega$.
As a result,
$$v_{y\backslash x}=|B(y,r)\cap \Omega|-|B(x,r)\cap B(y,r)\cap \Omega|\geq \frac{S}{2},$$
and therefore,
$$nv_{y\backslash x}\geq\frac{1}{2}nS\geq\frac{1}{2\sqrt{\pi}}\left(n\pi r^2\right)^{\frac12},$$
where  $S$ is given by Lemma~\ref{lem:shadow-low-bound}.
}
\begin{figure}[htbp]
 \begin{center}
       \scalebox{0.7}[0.7]{
           \includegraphics{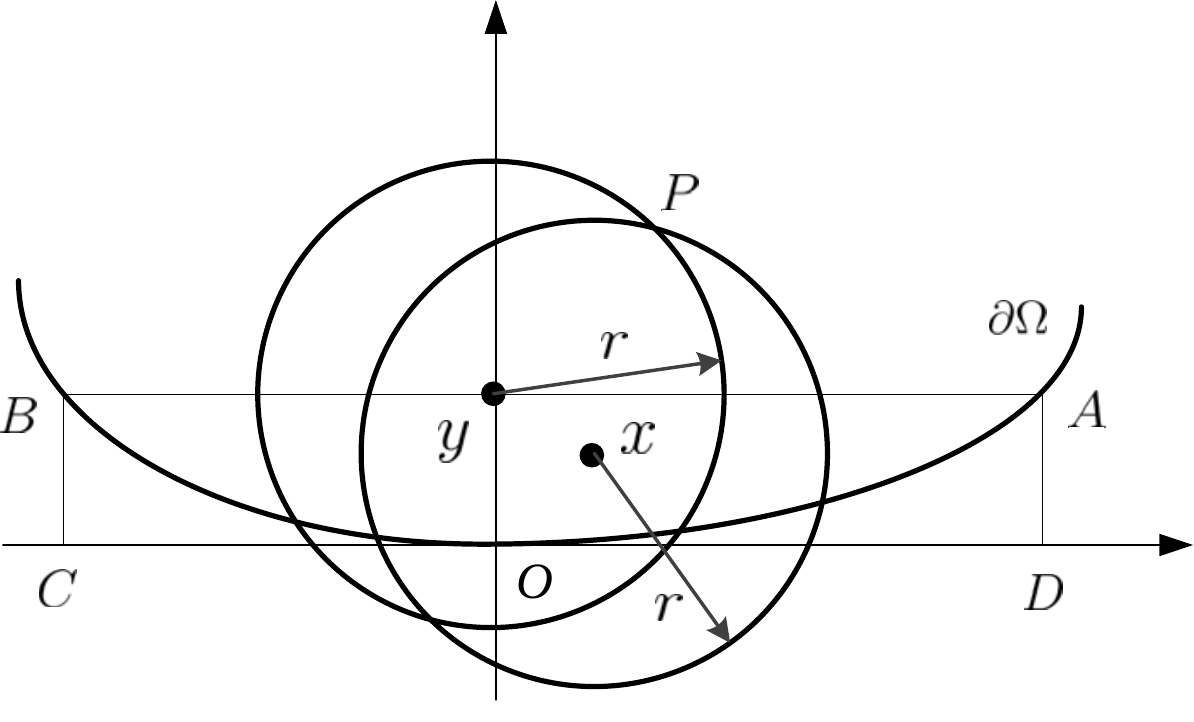} }
       \caption{Illustration of the proof for $A_2=o(1)$}\label{fig:lemma2-Application}
  \end{center}
  \end{figure}

So we have
 \begin{eqnarray*}
   &&A_2\\
   &=&\frac{1}{n\pi r^2}\int_{\Omega}n\psi^k_{n,r}(x)dx\int_{\Gamma_2}n\Pr(Z_3=k-1)dy\\
   &\leq&\frac{1}{n\pi r^2}\int_{\Omega}n\psi^k_{n,r}(x)dx\\
   &&\int_{\Gamma_2}\frac{n(n\pi r^2)^{k-1}\exp\left(-\frac C2\left(n\pi r^2\right)^{\frac12} \right) dy}{(k-1)!}\\
   &\leq&\frac{1}{n\pi r^2}\int_{\Omega}n\psi^k_{n,r}(x)dx
   \frac{(n\pi r^2)^{k}\exp\left(-\frac C2\left(n\pi r^2\right)^{\frac12} \right)}{(k-1)!}\\
   &\sim&\frac{e^{-c}}{n\pi r^2}o(1)\\
   &=&o(1).
 \end{eqnarray*}

\par Step 3: to prove $A_3=o(1)$ where $A_1$ is defined by (\ref{eq:A123}).
\begin{eqnarray*}
    A_3&=&\frac{1}{n\pi r^2}\int_{\Omega}n\psi^k_{n,r}(x)dx\int_{\Gamma_3(x)}n\Pr(Z_3=k-1)dy\\
      &\leq &\frac{1}{n\pi r^2}\int_{\Omega}n\psi^k_{n,r}(x)dx\int_{\Gamma_3(x)}ndy\\
      &=& \frac{1}{\pi r^2}\int_{\Omega}|\Gamma_3(x)|n\psi^k_{n,r}(x)dx\\
       &\leq& \frac{2\pi(G(\Omega)+1)r^3}{\pi r^2}\int_{\Omega}n\psi^k_{n,r}(x)dx\\
       &\sim&\Theta(r)\\
       &=&o(1).
\end{eqnarray*}
Finally, we have
\begin{equation*}
\begin{split}
&\frac{1}{n\pi r^2}\int_{\Omega}n\psi^k_{n,r}(x)dx\int_{\Omega\cap
B(x,r)}n\Pr(Z_3=k-1)dy\\
\leq &A_1+A_2+A_3=o(1).
\end{split}
\end{equation*}
The proposition is therefore proved.
\end{proof}

 \setcounter{lemma}{4}
 \setcounter{conclusion}{1}
\section{Proof Sketch of Conclusion~2}
{
\begin{conclusion} Under the assumptions of Theorem~\ref{thm:combined-Main},
$$\lim_{n\rightarrow\infty}\Pr\left\{\rho(\chi_n;\delta\geq k+1)=\rho(\chi_n;\kappa\geq
k+1)\right\}=1.$$
\end{conclusion}

Here we sketch the proof of Conclusion~2. Penrose has clearly proved this result when region $\Omega$ is a square~\cite{Penrose-k-connectivity}.  He cleverly constructed two events, $E_n(K)$ and $F_n(K)$,  such that for any $K>0$,
$$
     \left\{\rho(\chi_n;\delta\geq k+1)\leq r_n<\rho(\chi_n;\kappa\geq
k+1)\right\}\subseteq E_n(K)\cup F_n(K),
$$
and
$$
\lim_{n\rightarrow \infty}\Pr[E_n(K)]=\lim_{n\rightarrow \infty}\Pr[F_n(K)].
$$
The definition of the events and the convergence results are organised in
the following two propositions (see Proposition~5.1 and~5.2 in~\cite{Penrose-k-connectivity}).
Here and in the following, we do not introduce the notations in detail. Please refer to~\cite{Penrose-k-connectivity}.

\begin{proposition}(~\cite{Penrose-k-connectivity})\label{proposition-appendix-1}
Let $E_n(K)$ be the event that there exists a set $U$ of two or more
points of $\chi_n(U_{r_n})$, of diameter at most $Kr_n$ with
$\chi_n(U_{r_n}\backslash U)\leq k$. Then for all $K>0$ we have
$\lim_{n\rightarrow \infty}\Pr[E_n(K)]=0$.
\end{proposition}

\begin{proposition}\label{proposition-appendix-2}(~\cite{Penrose-k-connectivity})
Let $F_n(K)$ be the event that there are disjoint subsets $U,V,W$ of
$\chi_n$, such that $\mbox{card}(W)\leq k$, and such that $U$ and
$V$ are connected components of the $r_n$ graph on $\chi_n\setminus
W$, with $\mbox{diam}(U)>Kr_n$ and $\mbox{diam}(V)>Kr_n$. Then there
exists $K>0$, such that $\lim_{n\rightarrow \infty}\Pr[F_n(K)]=0$.
\end{proposition}

Proposition~\ref{proposition-appendix-1} is immediate from the following three lemmas (see Lemma~5.1, 5.2 and~5.3 in~\cite{Penrose-k-connectivity}).
\begin{lemma}(~\cite{Penrose-k-connectivity})
     Let $K>0$. Then with $\psi^{k}_{n,r}(x)$ defined previously,
     $$
          \Pr[E_n(K)]\leq \sup_{x\in\Omega} \left(
          \frac{\Pr[E_n^x(K)]}{\psi^{k}_{n,r_n}(x)}
          \right).
     $$
\end{lemma}

\begin{lemma}\label{lemma5}(~\cite{Penrose-k-connectivity}) There exists $K\in (0,1]$, such that
   $$
        \lim_{n\rightarrow\infty}\mathop{\sup}_{x\in \Omega}
        \frac{\Pr\left[E_n^x(K)\right]}{\psi_{n,r_n}^k(x)}=0.
   $$
\end{lemma}

\begin{lemma} (~\cite{Penrose-k-connectivity}) Suppose $0<K'<K<\infty$. Then
   $$
        \lim_{n\rightarrow\infty}\mathop{\sup}_{x\in \Omega}
        \frac{\Pr\left[E_n^x(K)\backslash E_n^x(K')\right]}{\psi_{n,r_n}^k(x)}=0.
   $$
\end{lemma}

These lemmas deal with square regions. However, it can be  straightly generalised  to cope with the convex case, with
their proofs not much been modified. That is, $|B(x,r)\cap\Omega|$ in  $
       \psi^k_{n,r}(x)= \frac{\left(n|B(x,r)\cap\Omega|\right)^k
       e^{-n|B(x,r)\cap\Omega|}}{k!}
$
can be bounded by $C\pi r^2\leq|B(x,r)\cap\Omega|\leq \pi r^2$ where $C>0$.
Estimation $n\int_{\Omega} \psi^k_{n,r}(x)dx\sim e^{-c}$ (see~(\ref{eq:proof-formulae-2})) can also hold.

As a result, Proposition~\ref{proposition-appendix-1} and~\ref{proposition-appendix-2} can be generalised to the case of
convex region.  Based on these two generalised propositions, a squeezing argument can lead to the following
$$\lim_{n\rightarrow\infty}\Pr\left\{\rho(\chi_n;\delta\geq k+1)=\rho(\chi_n;\kappa\geq
k+1)\right\}=1.$$
Please see the details presented in~\cite{Penrose-k-connectivity}.

}

\bibliographystyle{IEEEtran}
\bibliography{RGG-2021-Dec18-arxiv}

\end{document}